\documentclass[sigconf]{acmart}

\setcopyright{none}
\renewcommand\footnotetextcopyrightpermission[1]{} 
\AtBeginDocument{%
  }

\usepackage{nicefrac}
\usepackage{siunitx}
\usepackage{array,framed}
\usepackage{booktabs}
\usepackage{
  color,
  float,
  epsfig,
  wrapfig,
  graphics,
  graphicx,
  subcaption
}

\usepackage{setspace}
\usepackage{latexsym,fancyhdr,url}
\usepackage{enumerate}
\usepackage[ruled,vlined,linesnumbered]{algorithm2e}
\usepackage{algpseudocode}
\usepackage{xparse} 
\usepackage{xspace}
\usepackage{multirow}
\usepackage{csvsimple}
\usepackage{balance}

\usepackage{
  tikz,
  pgfplots,
  pgfplotstable
}
\usepackage{hyperref}

\usetikzlibrary{
  shapes.geometric,
  arrows,
  external,
  pgfplots.groupplots,
  matrix
}
\usepackage{mathtools,}

\usepackage{tabularx}
\usepackage{filecontents}
\usepackage{enumitem}

\begin{document}

\title{S-Leak: Leakage-Abuse Attack Against Efficient Conjunctive SSE via \texttt{s-term} Leakage}


\author{Yue Su}
\email{su_yue@bit.edu.cn}
\affiliation{%
  \institution{Beijing Institute of Technology}
  \city{Beijing}
  \country{China}
}

\author{Meng Shen}
\email{shenmeng@bit.edu.cn}
\authornote{Corresponding author}
\affiliation{%
  \institution{Beijing Institute of Technology}
  \city{Beijing}
  \country{China}
}

\author{Cong Zuo}
\email{zuocong10@gmail.com}
\affiliation{%
  \institution{Beijing Institute of Technology}
  \city{Beijing}
  \country{China}
}

\author{Yuzhi Liu}
\email{liuyuzhi@bit.edu.cn}
\affiliation{%
  \institution{Beijing Institute of Technology}
  \city{Beijing}
  \country{China}
}

\author{Liehuang Zhu}
\email{liehuangz@bit.edu.cn}
\affiliation{%
  \institution{Beijing Institute of Technology}
  \city{Beijing}
  \country{China}
}

\begin{abstract}
Conjunctive Searchable Symmetric Encryption (CSSE) enables secure conjunctive searches over encrypted data. While leakage-abuse attacks (LAAs) against single-keyword SSE have been extensively studied, their extension to conjunctive queries faces a critical challenge: the combinatorial explosion of candidate keyword combinations, leading to enormous time and space overhead for attacks. In this paper, we reveal a fundamental vulnerability in state-of-the-art CSSE schemes: \texttt{s-term} leakage, where the keyword with the minimal document frequency in a query leaks distinct patterns. We propose {\sffamily S-Leak}, the first passive attack framework that progressively recovers conjunctive queries by exploiting \texttt{s-term} leakage and global leakage. Our key innovation lies in a three-stage approach: identifying the \texttt{s-term} of queries, pruning low-probability keyword conjunctions, and reconstructing full queries. We propose novel metrics to better assess attacks in conjunctive query scenarios. Empirical evaluations on real-world datasets demonstrate that our attack is effective in diverse CSSE configurations. When considering 161,700 conjunctive keyword queries, our attack achieves a 95.15\% accuracy in recovering at least one keyword, 82.57\% for at least two, 58\% for all three keywords, and maintains efficacy against defenses such as SEAL padding and CLRZ obfuscation. Our work exposes the underestimated risks of \texttt{s-term} leakage in practical SSE deployments and calls for a redesign of leakage models for multi-keyword search scenarios. 
\end{abstract}



\keywords{Searchable Symmetric Encryption, Conjunctive Queries, Leakage-Abuse Attack}


\maketitle

\section{Introduction}
\label{sec:intro}
Searchable Symmetric Encryption (SSE)~\cite{sp00-song} schemes enable users to securely outsource datasets to cloud servers, while being able to perform secure queries over encrypted datasets. 
Conjunctive SSE (CSSE)~\cite{crypto13-oxt, ccs18-hxt, ndss21-odxt, ccs24-doris} is designed to enable secure search over conjunctive queries, an essential capability given that single-keyword queries are relatively rare in practice. Statistics indicate that the number of online searches peaks at two keywords~\cite{ckws-5}, with three-keyword queries still more frequent than single-keyword queries. 

However, most efficient SSE schemes~\cite{ccs16-bost, ccs18-hxt, ccs12-dynamic, sp14-blind} have been shown to be vulnerable to leakage-abuse attacks (LAAs)~\cite{ndss12-ikk, usenix16-all, ccs23-dynamic, usenix24-jigsaw}, where an honest-but-curious server could exploit leakage patterns and auxiliary information to recover the underlying keywords of client's queries or reconstruct the dataset. Numerous LAAs have been proposed over the past decade, most focus exclusively on single-keyword queries. 

Multi-keyword conjunctive queries introduce new challenges for LAAs, rendering direct adaptations of LAAs designed for single-keyword queries ineffective. \textit{Firstly}, candidate keyword conjunctions exhibit combinatorial growth with the number of keywords, increasing the attack complexity from $O(n)$ to $O(n^d)$, where $n$ is the number of keywords and $d$ is the maximum dimension of conjunctive queries. 
\textit{Secondly}, by returning only documents containing all queried keywords---as opposed to those matching individual keywords---the server inherently reduces the output dataset size. This diminished result volume lowers the entropy between distinct patterns, thereby increasing the attacker's uncertainty when attempting to reconstruct specific conjunctive queries. 
Existing study~\cite{usenix16-all} proposed active file injection attacks targeting conjunctive search schemes via Keyword Pair Result Pattern (KPRP), yet this method exhibits two inherent flaws: (1) the impracticality of server file injection under operational constraints, and (2) failure to address leakage resilience improvements demonstrated in modern KPRP optimizations~\cite{ccs18-hxt}. 
In this paper, we focus on the passive query recovery attack, which remains an open problem.

To address the aforementioned problems, we propose {\sffamily S-Leak}, the first passive attack framework against CSSE via \texttt{s-term} leakage. By systematically analyzing state-of-the-art CSSE schemes, we identify that most schemes are based on the OXT~\cite{crypto13-oxt} framework with a common query architecture involving the \texttt{s-term}, which has minimal document frequency in a conjunctive query. 
Except for the existing volume and equality pattern, we discover a novel \texttt{s-term} combination pattern based on the definition of \texttt{s-term}, which reveals the number of distinct queries sharing the same \texttt{s-term} within a query sequence. Leveraging these three \texttt{s-term} leakage patterns, we first recover all \texttt{s-term}s in the queries. 
Subsequently, we utilize the recovered \texttt{s-term}s to facilitate the reconstruction of full queries. Inspired by real-world query correlations similar to those analyzed in~\cite{usenix22-ihop}, we observe that keywords within conjunctive queries exhibit non-uniform co-occurrence patterns. Many keyword conjunctions demonstrate sufficiently weak correlations to be safely pruned from the vast conjunction space. Therefore, we can leverage the correlations between the recovered \texttt{s-term}s and other keywords to further recover the complete queries. 

Given the above inspiration, we propose the attack consisting of three core modules: (1) \textit{SRecover} identifies the \texttt{s-term} for each query using three leakage patterns: volume pattern, equality pattern and combination pattern with auxiliary information. We further group all queries by their \texttt{s-term} token for subsequent attack processes. (2) \textit{CandiPrun} prunes low-probability keyword combinations for each \texttt{s-term} by conjunctive query frequency analysis, and drastically reduces candidate keyword conjunctions which will be used in the next module. (3) \textit{FullRecover} reconstructs the remaining keywords for each query using the global access pattern and the global search pattern, while the candidate keyword conjunctions are refined with the output from \textit{CandiPrun}. We note that after completing the first module, the query list is grouped by \texttt{s-term}, enabling batch processing for subsequent modules and optimizing computational efficiency. This progressive approach mitigates combinatorial explosion while exploiting real-world query correlations to prune weakly-associated keyword pairs, significantly reducing both complexity and runtime compared to prior methods. 

Our main contributions are summarized as follows. 
\begin{itemize}
\item[$\bullet$] \textit{Leakage patterns analysis of efficient CSSE}: We review recent efficient conjunctive keyword search schemes and analyze their search processes. We first introduce \texttt{s-term} volume and equality pattern to LAAs, and identify a new \texttt{s-term} combination pattern that reveals the co-occurrence relationships with other keywords. 
\item[$\bullet$] \textit{Progressive attack framework for conjunctive queries}: We propose {\sffamily S-Leak}, the first passive LAA framework that progressively recovers conjunctive queries by exploiting s-term leakage and global leakage, as well as auxiliary datasets and novel query distribution. 
\item[$\bullet$] \textit{Practical Breakthrough in Attack Efficiency}: By exploiting the correlations among keywords in conjunctive queries, we address the challenge of combinatorial explosion, transforming the exponential $O(n^d)$ search space into a manageable $O(k\cdot n)$, where $k$ is the average number of candidates after pruning, making full query reconstruction feasible even for high-dimensional conjunctive search schemes. 
\item[$\bullet$] \textit{New Metrics and comprehensive performance evaluation}: We propose new metrics to better assess attacks in conjunctive query scenarios, among which the cumulative accuracy distribution ($CAD$), represents the accuracy of recovering at least $x$ keywords. Empirical experiments highlight the performance of {\sffamily S-Leak} in various CSSE settings on Enron and Lucene datasets. When considering 161,700 conjunctive keyword queries, our attack achieves a 95.15\% success rate in recovering at least one keyword, 82.57\% for at least two, 58\% for all three keywords, and maintains efficacy against defenses such as SEAL padding and CLRZ obfuscation.
\end{itemize}

\section{Preliminaries}
\label{sec:preli}
In this section, we introduce the background of CSSE and the leakage profiles of schemes based on the OXT~\cite{crypto13-oxt} framework.
\subsection{Background of CSSE}
Throughout this paper, we consider a two-party model, where the client owns a privacy-sensitive dataset that he/she intends to store on the remote server. The honest-but-curious server provides storage services, which faithfully executes the protocol while attempting to observe as much information as possible. To protect the dataset, the client encrypts each document using symmetric encryption and sends the encrypted documents to the server. Each document is associated with a set of keywords, and the client requires the ability to perform keyword-based searches.\footnote{It is worth noting that certain studies focus on datasets where each document is linked to a single keyword~\cite{usenix20-seal, ccs19-volhiding} (eg., the keyword may represent the document's publication date). We focus on attacks targeting schemes for conjunctive keyword queries, which require that at least some documents are associated with more than one keyword.} A Searchable Symmetric Encryption scheme SSE=(Setup, Search) that contains an algorithm and a protocol executed between the client and the server. The specific proceeding is described as follows.
\begin{itemize} 
\item[$\bullet$] Setup($\lambda$,\texttt{DB}) $\rightarrow$ (\texttt{K}, $\sigma$,\texttt{EDB}): The algorithm takes a database $\texttt{DB}={\{ind_i,\textbf{W}_i\}}$, where $ind_i$ is the file identifier, $\textbf{W}_i$ represents all keywords in the file $ind_i$, and a security parameter $\lambda$ as inputs, and outputs a secret key \texttt{K}, a local state $\sigma$ for the client, and an encrypted database \texttt{EDB} for the server.
\item[$\bullet$] Search(\texttt{K},$\sigma$,$q$,\texttt{EDB})$\rightarrow$(\texttt{R},$\perp$): The protocol runs between the client with the key \texttt{K}, the local state $\sigma$, and a query $q$ as inputs, and the server, which holds the \texttt{EDB}. Upon completion of the protocol, the client receives a set of files \texttt{R} that match to the query $q$, while the server receives no information. In this paper, we consider conjunctive keyword search schemes that support hybrid queries (i.e. if they support searching for conjunctions of up to three keywords, they also support searching for queries containing one or two keywords at the same time).
\end{itemize}
An SSE scheme is perfectly correct if the scheme retrieves all files matching the query. Our work focus on the schemes that ensure perfect correctness. 

\begin{figure}[!t]
    \centering 
    \includegraphics[width=0.47\textwidth]{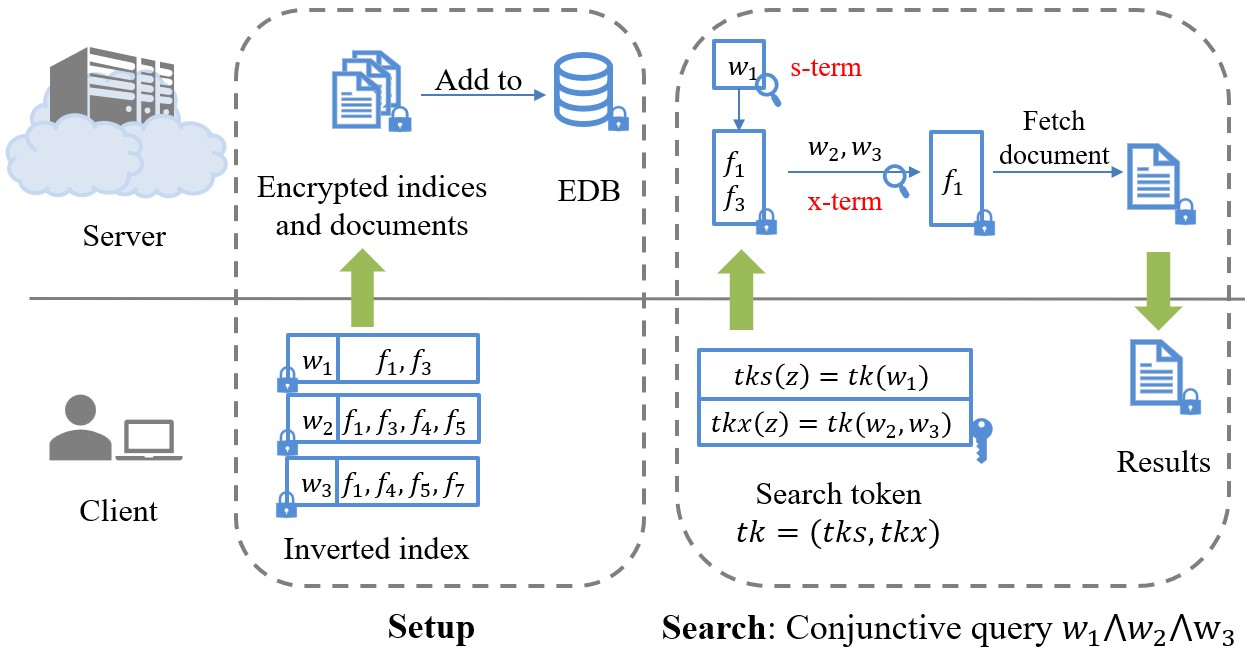} 
    \caption{Search process of representative CSSE schemes} 
    \label{fig:search_overview} 
\end{figure}
In the construction of conjunctive keyword search schemes, the specific design details of the search protocol can vary. However, modern implementations of conjunctive SSE~\cite{crypto13-oxt, ccs18-hxt, ccs24-doris} predominantly build upon the OXT~\cite{crypto13-oxt} framework. Specifically, they minimize search overhead by first querying the least frequent keyword in a conjunctive query, termed the \texttt{s-term}, defined as the keyword matching the fewest documents when queried individually. 
As illustrated in Figure~\ref{fig:search_overview}, the client generates a query token containing two parts: one for the \texttt{s-term}, the keyword with the least document frequency in the query, and another for the \texttt{x-term}, the remaining keywords. Then, upon receiving the query token, the server uses the \texttt{s-term} token to retrieve matching document identifiers. The server further filters these documents by using the other token to identify the subset of documents that match all keywords in the query, and corresponding encrypted documents can be retrieved in the final step. 
\subsection{Leakage Profiles}
\label{sec:pre_leakage}
An efficient CSSE scheme typically incurs both global and \texttt{s-term} leakage as we illustrated above. Global leakage refers to the information leaked about the entire conjunctive query, which can be divided into the global access pattern and the global search pattern. On the other hand, \texttt{s-term} leakage pertains to the leakage specific to the \texttt{s-term} within the conjunctive query, and can be divided into the \texttt{s-term} volume pattern, the \texttt{s-term} equality pattern and the \texttt{s-term} combination pattern. In particular, in schemes~\cite{crypto13-oxt, ccs18-hxt, ndss21-odxt, ccs24-doris} that rely on the OXT framework, only the size of the document set matching the \texttt{s-term} is leaked during the search process, and the actual document identifiers are not revealed. We focus on the minimal leakage information in schemes that involve the \texttt{s-term}, although some schemes reveal additional information beyond this basic leakage model. 

It is worth emphasizing that we are the first to formally characterize these leakage patterns from the attacker’s perspective. Among them, although the \texttt{s-term} volume pattern and the \texttt{s-term} equality pattern have been mentioned in some CSSE schemes, we have made their applications in the attack model more precise through more comprehensive analysis and formal definitions. On the other hand, the \texttt{s-term} combination pattern is a novel \texttt{s-term} leakage pattern proposed by us. It is derived from the volume correlation among conjunctive keywords. We will present our relevant observations in Section \ref{sec:method_1}. For a sequence of $\rho$ queries $Q^{\rho}=[q_1=({q_1}_s, {q_1}_x), \ldots ,q_{\rho}=({q_\rho}_s, {q_\rho}_x)]$, all leakage patterns we considered are summarized as follows.
\begin{itemize} 
\item[$\bullet$] \textbf{Global access pattern} denoted as $AP=[ID(q_1),\ldots,\\ID(q_{\rho})]$, where $ID(\cdot)$ represents the document identifiers that match the entire query token. For each query, the scheme leaks the identifiers of encrypted documents that match all keywords in the conjunction. This leakage occurs in all CSSE schemes when the user retrieves the documents.
\item[$\bullet$] \textbf{Global search pattern} denoted as a ${\rho}\times {\rho}$ binary matrix $QEQ^{{\rho}\times {\rho}}$, where $QEQ^{{\rho}\times {\rho}}[i,j]=1$, if the underlying conjunctive keywords of $q_i$ and $q_j$ are completely identical and otherwise $QEQ^{{\rho}\times {\rho}}[i,j]=0$. For any two queries $q_i, q_j \in Q^{\rho}, i \neq j$, the attacker knows whether $q_i$ has the same underlying keywords as $q_j$. 
\item[$\bullet$] \textbf{\texttt{s-term} volume pattern} denoted as $SVOL=[|ID({q_1}_s)|,\ldots,\\|ID({q_\rho}_s)|]$, where $|\cdot|$ represents cardinality of the elements inside. For each query, the scheme leaks the number of documents that match the \texttt{s-term} of the conjunctive query.
\item[$\bullet$] \textbf{\texttt{s-term} equality pattern} denoted as a ${\rho}\times {\rho}$ binary matrix $SEQ^{{\rho}\times {\rho}}$, where $SEQ^{{\rho}\times {\rho}}[i,j]=1$, if the underlying \texttt{s-term} of query $q_i$ and $q_j$ is the same and otherwise $SEQ^{{\rho}\times {\rho}}[i,j]=0$. For any two queries $q_i, q_j \in Q^{\rho}, i \neq j$, the attacker knows whether ${q_i}$ has the same underlying \texttt{s-term} as ${q_j}$.
\item[$\bullet$] \textbf{\texttt{s-term} combination pattern} denoted as $SCN=[m_1,\ldots,\\ m_{n_s}]$, where $n_s$ is the number of distinct \texttt{s-term} tokens in the $Q^{\rho}$ and $m_i$ denotes the number of distinct query with the same $i$-th \texttt{s-term}. $SCN$ represents how many distinct query tokens related to the \texttt{s-term}. To further illustrate SCN, we consider a set of queries $\{(\underline{w_1},w_2), (\underline{w_1},w_3), (\underline{w_1},w_3), (\underline{w_2},w_3), \\ (\underline{w_2},w_4), (\underline{w_2},w_4), (\underline{w_2},w_5), (\underline{w_3},w_4)\}$, where the keyword with an underline denotes the \texttt{s-term} of the query. In this example, $n_s = 3$ and $SCN = [2, 3, 1]$.
\end{itemize}
\section{Attack Model}
\label{sec:attackmodel}
We focus on passive attack and consider an honest-but-curious server as the attacker. The server has full access to the encrypted documents and follows the CSSE protocols and always returns the correct result for each query, but tries to learn as much information as possible. In this paper, the attacker's goal is to perform a query recovery attack, aiming to identify the underlying keywords associated with each query token. The attack result is an injective mapping from the set of query tokens to the set of keyword conjunctions. Notations that we used are summarized in Table~\ref{tab:notation}.
\begin{table}[h]
\caption{Summary of notations.} 
\small
\centering
\begin{tabularx}{0.48\textwidth}{|c|X|}
\hline
\multicolumn{2}{|c|}{\textbf{Auxiliary (Background) Information}} \\ \hline
$\Delta_k$                    & Keyword universe $\Delta_k=[w_1, w_2, \ldots, w_n]$ \\ \hline
$\Delta_{c}$                  & Keyword combination universe, $\Delta_{c}=[z_1, z_2, \ldots, z_{n_c}]$ \\ \hline
${\Delta^i_c}$                & Keyword combination universe with \texttt{s-term} $w_i$, ${\Delta^i_c}=[z^{i}_1, z^{i}_2, \ldots, z^{i}_{\tilde{m}_i}]$ \\ \hline
${\Delta^i_c}^{'}$            & Filtered candidate keyword combination universe with \texttt{s-term} $w_i$, ${\Delta^i_c}^{'}=[z^i_1, z^i_2, \ldots, z^i_{\tilde{m}_i \beta_i}]$ \\ \hline
$\tilde{\textbf{m}}$          & Number vector of keyword combinations with the same \texttt{s-term}, $\tilde{\textbf{m}}=[\tilde{m}_1, \tilde{m}_2, \ldots, \tilde{m}_n]$, L1-normalized version is denoted by~$\tilde{\textbf{m}}^*$\\ \hline
$\tilde{\textbf{v}}$          & Volume vector of keywords, $\tilde{\textbf{v}}=[\tilde{v}_1, \tilde{v}_2, \ldots, \tilde{v}_n]$ \\ \hline
$\widetilde{\textbf{Sf}}$     & \texttt{s-term} query frequency vector of all keywords, $\widetilde{\textbf{Sf}}=[\widetilde{Sf}_1, \widetilde{Sf}_2, \ldots, \widetilde{Sf}_n]$ \\ \hline
$\widetilde{\textbf{V}_i}^{'}$ & Volume matrix of keyword combinations in ${\Delta^i_c}^{'}$ \\ \hline
$\widetilde{\textbf{f}_i}^{'}$ & Query frequency vector of keyword combinations in ${\Delta^i_c}^{'}$, $\widetilde{\textbf{f}_i}^{'}=[\widetilde{f_i}_1, \widetilde{f_i}_2, \ldots, \widetilde{f_i}_{\tilde{m}_i \beta}]$ \\ \hline
\multicolumn{2}{|c|}{\textbf{Attacker Observations}} \\ \hline
$\Delta_\gamma$               & Query \texttt{s-term} token universe, $\Delta_\gamma=[\gamma_1, \gamma_2, \ldots, \gamma_{n_s}]$ \\ \hline
${\Delta^u_{\tau}}$           & Universe of query token whose \texttt{s-term} token is $\gamma_u$, ${\Delta^u_{\tau}}=[\tau^u_1, \tau^u_2, \ldots, \tau^u_{m_u}]$ \\ \hline
$\textbf{m}$                  & Number vector of distinct query tokens with the same \texttt{s-term} token, $\textbf{m}=[m_1, m_2, \ldots, m_{n_s}]$, L1-normalized version is denoted by~$\textbf{m}^*$ \\ \hline
$\textbf{v}$                  & Volume pattern of \texttt{s-term} tokens, $\textbf{v}=[v_1, v_2, \ldots, v_{n_s}]$ \\ \hline
$\textbf{Sf}$                 & \texttt{s-term} query frequency vector of all \texttt{s-term} tokens, $\textbf{Sf}=[Sf_1, Sf_2, \ldots, Sf_{n_s}]$ \\ \hline
$\textbf{V}_u$                & Volume matrix of observed tokens whose \texttt{s-term} token is~$\gamma_u$ \\ \hline
$\textbf{f}_u$                & Query frequency vector of tokens whose \texttt{s-term} token is $\gamma_u$, $\textbf{f}_u=[{f_u}_1, {f_u}_2, \ldots, {f_u}_{{m}_u}]$ \\ \hline
\multicolumn{2}{|c|}{\textbf{General Parameters}} \\ \hline
$d$                           & Maximum dimension of conjunctive queries \\ \hline
$P_d$                         & Probability of query varying numbers of keywords in the hybrid query setting \\ \hline
$\rho$                        & Number of queries issued by the client \\ \hline
$N_D$                         & Number of documents in the encrypted dataset \\ \hline
\textbf{$\beta$}              & Scale factor \\ \hline
\end{tabularx}
\label{tab:notation} 
\vspace{-10pt}
\end{table}

\noindent \textbf{Attacker's knowledge derived from leakages.}
We use the leakage patterns to derive the observation of the attacker. Specifically, we consider the attacker's observation knowledge from two perspectives: \texttt{s-term} leakage and global leakage. The server is aware of the total number of encrypted documents, denoted as $N_D$. 

We assume the client generates $\rho$ queries, denoted as $Q^{\rho}$, from which the attacker can identify $n_s$ different \texttt{s-term} tokens by the \texttt{s-term} equality pattern. We denote the distinct \texttt{s-term} tokens as $\Delta_\gamma=[\gamma_1,\ldots,\gamma_{n_s}]$. For each \texttt{s-term} token $\gamma$, we normalize the \texttt{s-term} volume pattern of the $u$-th \texttt{s-term} token denoted as $v_u=|ID(\gamma_u)|/N_D$, where $u \in [n_s]$, and $\textbf{v}=[v_1,\ldots,v_{n_s}]$ is the volume. With the \texttt{s-term} equality pattern, the attacker can compute the frequency of $Sf_u=Count(\gamma_u)/{\rho}$, where $Count(\gamma_u)$ calculates the number of $\gamma_u$ as the \texttt{s-term} token of $Q^\rho$, and $\textbf{Sf}=[Sf_1,\ldots,Sf_{n_s}]$ is the frequency of the \texttt{s-term} tokens in $\Delta_\gamma$. 

We divide the entire query list according to the same \texttt{s-term} tokens, then we obtain $n_s$ query sublists. The sublist of the $u$-th \texttt{s-term} token can be denoted as $Q_u=[q^u_1, \ldots,q^u_{{\rho}_u}], u\in [n_s]$, where ${\rho}_u$ is the number of query tokens with the same \texttt{s-term} token $\gamma_u$. For the $u$-th query sublist, the attacker obtains the query token universe with the same \texttt{s-term} token $\gamma_u$ by further leveraging the global search pattern, which can be denoted as $\Delta^{u}_\tau=[\tau^{u}_1,\ldots,\tau^{u}_{m_u}]$, where $m_u$ is the number of distinct query tokens with \texttt{s-term} token $\gamma_u$. We normalize the \texttt{s-term} combination pattern of the $u$-th \texttt{s-term} token denoted as $m_u^*=m_u/n_c$, where $u \in [n_s]$ and $n_c$ denotes the number of all possible keyword conjunctions, and $\textbf{m}^*=[m_1^*,\ldots,m_{n_s}^*]$ is the normalized \texttt{s-term} combination number. With the global access pattern, the attacker acquires knowledge of the returned documents $[D(\tau^{u}_1),\ldots,D(\tau^{u}_{m_u})]$. The access pattern of token $\tau^{u}_j$, ${\textbf{a}_u}_j$ can be constructed as a $1 \times N_D$ vector, whose $i$-th entry is 1 if the $i$-th document of the dataset matches the query, and 0 otherwise. The matrix of observed volumes $\textbf{V}_u$ from access pattern is an $m_u\times m_u$ matrix whose $j,j'$-th entry represents the number of documents matching both query tokens $\tau^{u}_j$ and $\tau^{u}_{j'}$, i.e., $(\textbf{V}_{u})_{j,j'}={\textbf{a}_u}_{j'}\cdot {\textbf{a}_u}_j^T/N_D$. The observed query frequency, $\textbf{f}_u$, is of length $m_u$, where the $j$-th entry represents the number of times the client queried for $\tau^u_j$, normalized by the length of the sublist~${\rho}_u$.

\noindent \textbf{Attacker's knowledge derived from similar data. }
Similar to~\cite{usenix24-jigsaw,usenix22-ihop,usenix21-sap}, we assume the attacker possesses auxiliary information $D_a$ in the form of similar documents---an assumption that aligns with weaker known-data hypotheses (in contrast to stronger known-data assumptions adopted in~\cite{ccs21-leap,ccs15-count,ndss20-revisiting})---and employs the same keyword extraction algorithm as the client. 
We denote the keyword universe extracted by the attacker as $\Delta_k=[w_1,\ldots,w_n]$. Then the attacker can construct the single keyword volume $\tilde{\textbf{v}}=[\tilde{v}_1,\ldots,\tilde{v}_n]$, where $\tilde{v}_i=|D_a(w_i)|/|D_a|$, and $D_a(w_i)$ is the documents in $D_a$ containing keyword $w_i$. By computing the combinations of elements in the keyword universe, the keyword conjunction universe can be obtained as $\Delta_{c}=[z_1,\ldots,z_{n_c}]$. 

Based on the document frequency of keywords in the auxiliary dataset $D_a$, the keyword conjunctions in $\Delta_{c}$ can be divided according to the same \texttt{s-term}, and $n$ sub-universes can be obtained.\footnote{Note that after extracting \texttt{s-term}, we can get $n$ subsets instead of $n-1$ subsets. Because hybrid queries are considered, and the keyword with the highest document frequency may also be queried by the user as a single keyword.} The universe of keyword conjunctions with $w_i$ as the \texttt{s-term} can be expressed as ${\Delta^i_c}=[z^i_1,\ldots,z^i_{\tilde{m}_i}]$, where $\tilde{m}_i$ is the number of keyword conjunctions with the \texttt{s-term} $w_i$ and the normalized one can be denoted as $\tilde{\textbf{m}}^*$. 
The attacker also constructs access pattern $\widetilde{\textbf{a}_i}_g$ from $D_a$ in a similar way to the construction of ${\textbf{a}_u}_j$. Then the attacker computes the co-occurrence volume matrix with the same \texttt{s-term} $w_i$ as $\widetilde{\textbf{V}_i}$, whose $g,g'$-th entry represents the number of documents matching both keyword conjunctions $z^i_g$ and $z^i_{g'}$, i.e., $(\widetilde{\textbf{V}_i})_{g,g'}=\widetilde{\textbf{a}_i}_{g'} \cdot \widetilde{\textbf{a}_i}_g^T/|D_a|$. 

The attacker can also obtain query frequency by public information~\cite{usenix21-sap,usenix22-ihop,usenix24-jigsaw} such as Google Trend or outdated query frequency information. We assume the attacker can access query frequencies for both single keywords and 2-dimensional keyword conjunctions, with the latter representing a novel attack vector that has not been explored in prior research. Query frequencies obtained from public information do not directly provide the query frequency of keywords as \texttt{s-term}s. However, the attack can derive the \texttt{s-term} query frequencies of keywords $\widetilde{\textbf{Sf}} = [\widetilde{Sf}_1, \ldots, \widetilde{Sf}_n]$, along with the query frequencies when $w_i$ serves as the \texttt{s-term}, denoted as $\widetilde{\textbf{f}_i} = [\widetilde{f_i}_1,\ldots, \widetilde{f_i}_{\tilde{m}_i}]$. The detailed processing procedure is described in Appendix \ref{app:freq_process}.

\section{The proposed {\sffamily S-Leak}}
\label{sec:method}
In this section, we present the key observation underlying our attack design and the attack overview, laying the foundation for the subsequent design details.

\subsection{Key Observation}
\label{sec:method_1}
In this subsection, we first delineate the foundational observations and rationale behind our proposed \textit{\texttt{s-term} combination pattern (SCN)}, then demonstrate the observations on the correlations between keywords in conjunctive queries.

\noindent \textbf{\texttt{s-term} combination pattern (SCN). }
As mentioned in Section \ref{sec:preli}, the \texttt{s-term} is a crucial part of conjunctive query. During the entire search process, it is queried as a single keyword, leaking independent information. For the query recovery attack on conjunctive keywords, an obvious intuition is to first recover the \texttt{s-term}, and then recover the full query. Besides the commonly leaked volume pattern and equality pattern in single-keyword query processes, the definition of the \texttt{s-term} in a conjunctive query also provides a breakthrough for leakage abuse. The \texttt{s-term} is the keyword with the least document frequency among those involved in the conjunctive query. Evidently, a keyword with a lower document frequency is more likely to be the \texttt{s-term} of a conjunctive query. Therefore, we can also count the number of different keyword conjunctions in which these keywords serve as the \texttt{s-term}, and use this combination count as a pattern to participate in the \texttt{s-term} recovery process. This pattern is the \texttt{s-term} combination pattern (SCN) proposed by us, which reflects the relative volume relationship among the keywords participating in the conjunctive query. 
\begin{figure}[!t]
        \centering 
        \begin{subfigure}[b]{0.18\textwidth}
            \label{fig:scn_d2} 
            \includegraphics[height=0.8\textwidth, width=\textwidth]{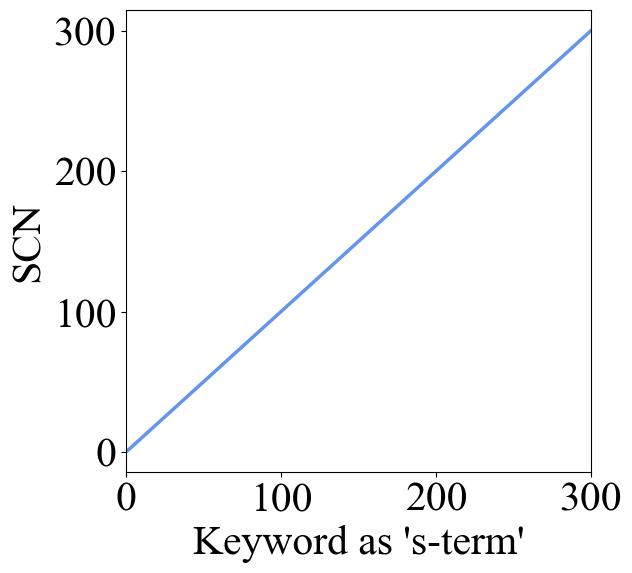}
            \caption{\normalfont d=2}
        \end{subfigure}
        \hspace{20pt}
        \begin{subfigure}[b]{0.18\textwidth}
            \label{fig:scn_d3} 
            \includegraphics[height=0.8\textwidth, width=\textwidth]{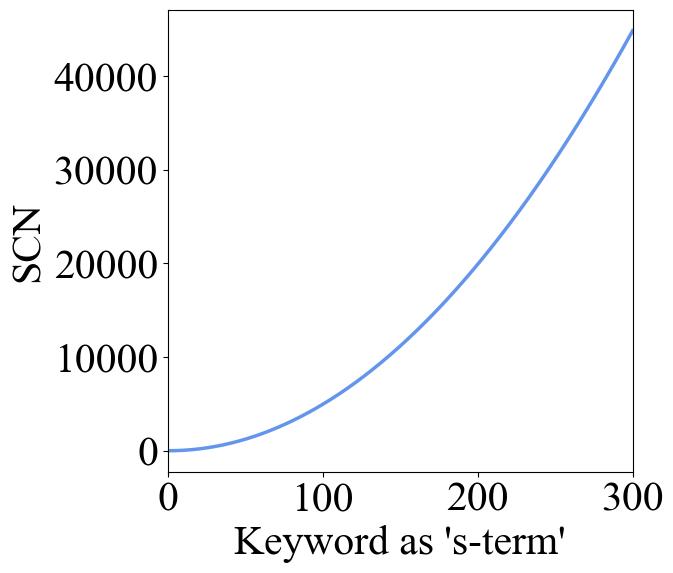} 
            \caption{\normalfont d=3}
        \end{subfigure}
        \caption{The \texttt{s-term} combination pattern of each keywords in hybrid query setting.} 
        \label{fig:scn} 
\end{figure}
\begin{figure}[!t]
    \centering
    \begin{subfigure}[t]{0.225\textwidth}
        \includegraphics[height=0.6\textwidth, width=\textwidth]{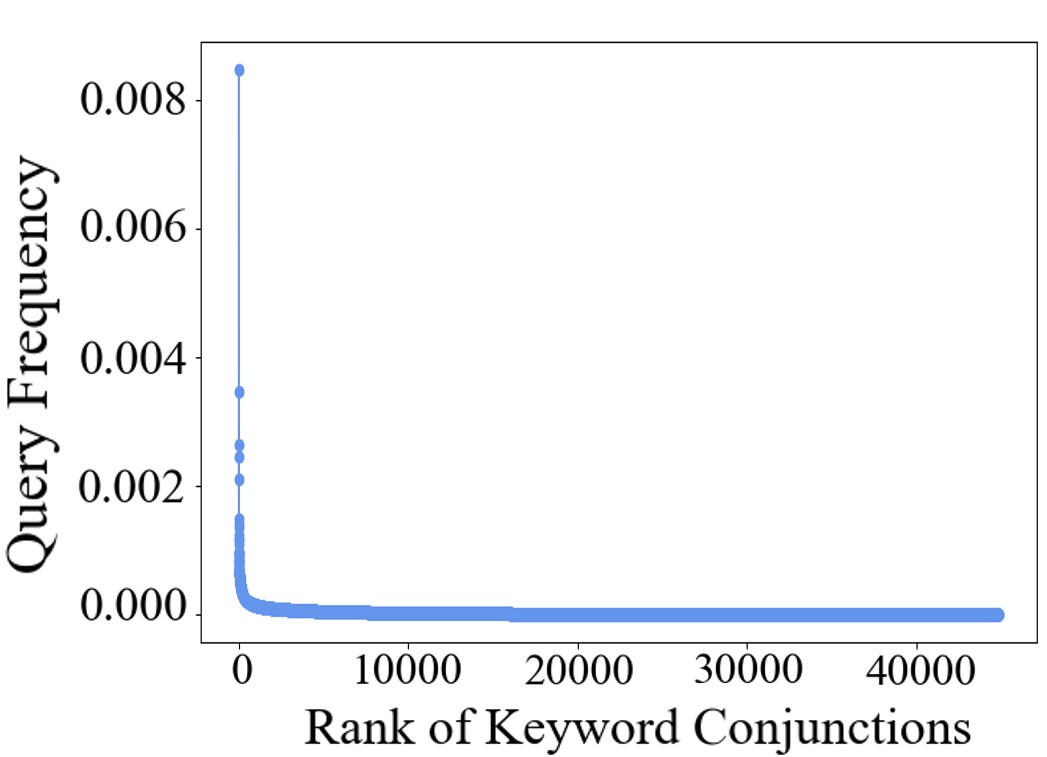}
        \caption{\normalfont Query frequency of conjunctive keywords.}
        \label{fig:m2obs1}
    \end{subfigure}
    \hspace{10pt}
    \begin{subfigure}[t]{0.223\textwidth}
        \includegraphics[height=0.6\textwidth, width=\textwidth]{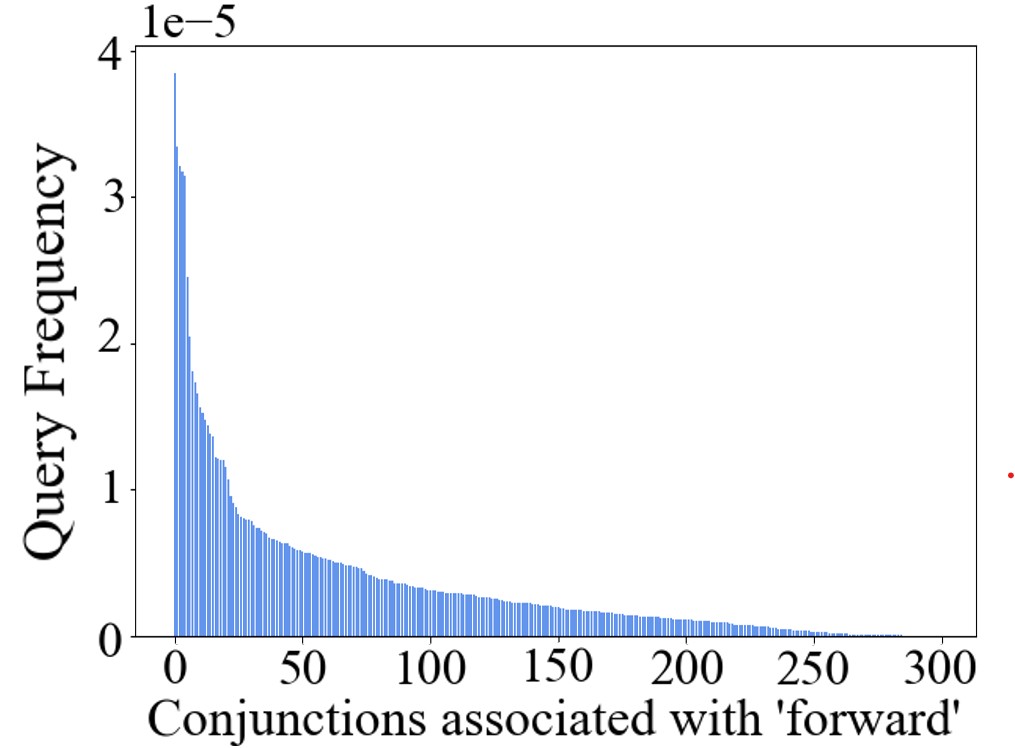}
        \caption{\normalfont Query frequency associated with `forward'.}
        \label{fig:m2obs2}
    \end{subfigure}%
    \caption{Conjunctive query frequency from Google Trends.}
\end{figure}

We considered all keyword combinations in hybrid queries with 300 keywords under $d=2$ and $d=3$, and visualized the SCN of each keyword as shown in Figure~\ref{fig:scn}. It indicates that different keywords exhibit distinct SCN values and this difference can be leveraged as a pattern in the \texttt{s-term} recovery process. 
Note that the sorted SCN results are the same, because we assume that all keyword combinations will be queried, and the volume relationship among the keywords is fixed.

\noindent \textbf{Correlations between keywords in conjunctive queries. }
In query recovery attacks targeting multi-keyword conjunctive search schemes, treating each keyword conjunction equally leads to an exponentially expanding search space ($O(n^d)$, where $n$ is the number of keywords and $d$ is the maximum dimension of conjunctive queries. ). This incurs excessive time costs, making attacks infeasible even for moderately sized keyword sets or schemes supporting higher-dimensional conjunctive queries. However, in real-world scenarios, keywords within conjunctive queries often exhibit correlations. Within the vast keyword conjunctions space, numerous low-correlation keyword conjunctions have negligible probabilities of being queried together.  

In this paper, we focus on the query frequencies of two-keyword conjunctive queries, which can reflect the correlation between keywords and approximately extend to the frequencies of queries involving more keywords by utilizing frequency estimation methods. The detailed estimation procedure is described in Appendix \ref{app:freq_approx}. 
We analyze the query frequencies of 2-dimensional conjunctive queries from the top 300 most frequent keyword stems in the Enron dataset. The visualized statistical results are shown in Figure \ref{fig:m2obs1}, which demonstrates a sharp frequency decline: the top-ranked conjunction has a query frequency of approximately 0.008, dropping sharply to near 0 over a small rank range, then remaining extremely low across a broad rank span. Additionally, we visualized the query frequencies of conjunctive queries containing the keyword `forward' as shown in Figure \ref{fig:m2obs2}. The frequency begins around \(4\times10^{-5}\) for the most frequent conjunction and decreases rapidly, forming a long-tailed distribution. 
It can be observed that for the same keyword, different keyword conjunctions demonstrate a significantly different conjunctive query frequency, and the distribution of these query frequencies follows the Zipf's law. 

Thus, once a keyword in the conjunctive query has been recovered, we can leverage this correlation to aid in the further recovery of the remaining keywords in the query. 
This correlation reflects real-world scenarios and directly supports our progressively recovery approach: first recover the \texttt{s-term}, then reconstruct the full query using pruned candidate sets.

\subsection{Overview}
\label{sec:method_2}
In response to the unresolved challenge of passive query recovery, we propose {\sffamily S-Leak}, the first framework addressing the combinatorial explosion problem in CSSE by exploiting \textit{\texttt{s-term} leakage} and \textit{global leakage patterns}. The core methodology of {\sffamily S-Leak} operates through a three-stage progressive recovery mechanism. 


\noindent \textbf{\texttt{s-term} recovery in conjunctive queries (\textit{SRecover}).} 
The deterministic \texttt{s-term} selection mechanism in state-of-the-art CSSE schemes creates additional observable leakage information. Based on the leakage pattern identified in Section \ref{sec:method_1}, this module integrates the \texttt{s-term} combination pattern together with the volume pattern and the equality pattern to recover the \texttt{s-term} of every conjunctive query. Thus, the recovered \texttt{s-term} can provide knowledge for subsequent keyword recovery.

\noindent \textbf{Candidate keyword conjunction pruning (\textit{CandiPrun}).} 
Having recovered the \texttt{s-term} of each query, we can partition all observed queries according to the \texttt{s-term}. For each partitioned group, we implement a threshold-based pruning strategy that leverages the correlations between each conjunctive keyword and the corresponding \texttt{s-term}, based on the observation. 
This reduces the complexity of the subsequent full query recovery and achieves a reduction in dimensionality for the combinatorial explosion problem. 

\noindent \textbf{Full query reconstruction (\textit{FullRecover}).} 
For each group partitioned in the \textit{CandiPrun}, we subsequently analyze the pruned candidate keyword conjunctions by correlating the global access pattern with the search pattern, ultimately reconstructing the full conjunctive queries.


\section{Design Details}
In this section, we present the design details of our {\sffamily S-Leak} attack.

\subsection{\texttt{s-term} Recovery in Conjunctive Queries}
In the first module \textit{SRecover}, we leverage the three \texttt{s-term} leakage patterns to construct a maximum likelihood mapping between the knowledge observed by the attacker and that derived from the auxiliary dataset, thereby recovering the \texttt{s-term} for each query.

We look for the mapping $\textbf{SP}$ that maximizes the likelihood of observing $\textbf{v}$, $\textbf{Sf}$, $\textbf{m}^*$, $\rho$, $n_{c}$ and $N_D$ given the auxiliary information $\tilde{\textbf{v}}$, $\widetilde{\textbf{Sf}}$ and $\tilde{\textbf{m}}^*$. Due to the large number of keyword conjunctions and the impracticality of querying all possible keyword conjunctions in most cases, we scale $\tilde{\textbf{m}}^*$ by \textbf{$\beta$}$=[\beta_1,\ldots,\beta_n]$ to represent the new $\tilde{\textbf{m}}^*$, which is then used for attack matching with $\textbf{m}^*$. The detailed rationale is explained in the second module \textit{CandiPrun}. Formally, our attack solves the maximum likelihood problem
\begin{equation}
\textbf{SP}=\underset{\textbf{SP}\in\mathcal{SP}}{\texttt{argmax}} \operatorname{Pr}(\textbf{Sf},\rho, \textbf{v}, N_D, \textbf{m}^*, n_{c}  \mid \widetilde{\textbf{Sf}},\tilde{\textbf{v}}, \tilde{\textbf{m}}^*,\textbf{SP}).
\label{m1:argmax}
\end{equation}
We aim at to characterize $\textbf{Sf},\rho, \textbf{v}, N_D, \textbf{m}^*$ and $n_{c}$ given $\widetilde{\textbf{Sf}},\tilde{\textbf{v}}, \tilde{\textbf{m}}^*$, and an assignment of \texttt{s-term} tags to keywords $\textbf{SP}$. We assume that the user's querying behavior, the response volume, and the \texttt{s-term} combination number are independent, i.e., 
\begin{equation}
\begin{aligned}
&\operatorname{Pr}(\textbf{Sf},\rho, \textbf{v}, N_D, \textbf{m}^*, n_{c}  \mid \widetilde{\textbf{Sf}},\tilde{\textbf{v}}, \tilde{\textbf{m}}^*,\textbf{SP}) \\=
&\operatorname{Pr}(\textbf{Sf},\rho \mid \widetilde{\textbf{Sf}},\textbf{SP}) \cdot \operatorname{Pr}(\textbf{v}, N_D \mid \tilde{\textbf{v}},\textbf{SP}) \cdot \operatorname{Pr}(\textbf{m}^*, n_{c} \mid  \tilde{\textbf{m}}^*,\textbf{SP}).
\end{aligned}
\label{m1:independent}
\end{equation}

In our model, the client chooses the conjunctive keywords of each query independently from other queries following the query frequencies. This also means that the number of queries whose \texttt{s-term} is the keyword $w_i$ follows a Poisson distribution with $\rho$ trials and probabilities given by $\widetilde{\textbf{Sf}}$. Formally,
\begin{equation}
\begin{aligned}
&\operatorname{Pr}(\textbf{Sf},\rho \mid \widetilde{\textbf{Sf}},\textbf{SP})=\operatorname{Pr}(\rho) \cdot \operatorname{Pr}(\textbf{Sf} \mid  \widetilde{\textbf{Sf}},\rho,\textbf{SP}) \\=
&\operatorname{Pr}(\rho) \cdot \prod\limits_{u=1}^s\frac{(\widetilde{Sf}_{sp(u)})^{\rho \cdot Sf_{u}}}{(\rho \cdot Sf_{u})!}.
\end{aligned}
\label{m1:Prsf}
\end{equation}

The total document number $N_D$ in the encrypted database is independent of mapping $\textbf{SP}$, and keywords in each encrypted document are independently selected. Given the relative keyword volumes $\tilde{\textbf{v}}=[\tilde{v}_1, \ldots, \tilde{v}_n]$ from auxiliary information, each document is assigned to keyword $w_i$ with probability $\tilde{v}_i$. Thus, the number of documents returned for query $w_i$ follows a Binomial distribution with $N_D$ trials and success probability $\tilde{v}_i$. Formally,
\begin{equation}
\begin{aligned}
&\operatorname{Pr}(\textbf{v}, N_D \mid \tilde{\textbf{v}}, \textbf{SP})=\operatorname{Pr}(N_D) \cdot \operatorname{Pr}(\textbf{v} \mid \tilde{\textbf{v}}, N_D, \textbf{SP}) \\=
&\operatorname{Pr}(N_D) \cdot \prod\limits_{u=1}^s\binom{N_D}{N_D v_u} \widetilde{v}_{sp(u)}^{N_D v_u}(1-\widetilde{v}_{sp(u)})^{N_D(1-v_u)}.
\end{aligned}
\label{m1:Prv}
\end{equation}

Similar to the $\tilde{\textbf{v}}$, 
based on the scaled relative \texttt{s-term} combination numbers $\tilde{\textbf{m}}^*=[\tilde{m}_1^*,\ldots,\tilde{m}_n^*]$, each keyword conjunction has \texttt{s-term} $w_i$ with probability $\tilde{m}_i^*$. The \texttt{s-term} combination number for $w_i$ follows a Binomial distribution with $n_{c}$ trials and success probability $\tilde{m}_i^*$. Formally, 
\begin{equation}
\begin{aligned}
&\operatorname{Pr}(\textbf{m}^*, n_{c} \mid  \tilde{\textbf{m}}^*,\textbf{SP})=\operatorname{Pr}(n_{c}) \cdot \operatorname{Pr}(\textbf{m}^* \mid \tilde{\textbf{m}}^*, n_{c}, \textbf{SP}) \\=
&\operatorname{Pr}(n_{c}) \cdot \prod\limits_{u=1}^s\binom{n_{c}}{n_{c} m_u^*}
{\tilde{m}^{*^{n_c m^*_u}}_{sp(u)}}{(1-\tilde{m}^*_{sp(u)})}^{n_{c}(1-m_u^*)}.
\end{aligned}
\label{m1:Prm}
\end{equation}

We use maximum likelihood estimator to find $\textbf{SP}$ that maximizes $\operatorname{Pr}(\textbf{Sf},\rho, \textbf{v}, N_D, \textbf{m}^*, n_{c}  \mid \widetilde{\textbf{Sf}},\tilde{\textbf{v}}, \tilde{\textbf{m}}^*,\textbf{SP})$. We transform this optimization problem into minimizing the negative logarithm of this probability to avoid precision issues. The additive terms can be ignored in the objective function that are independent of $\textbf{SP}$, since they do not affect the optimization problem. Thus, the final log-likelihood cost of assigning $w_i\to\gamma_u$ is $(C_f)_{i,u}+(C_v)_{i,u}+(C_m)_{i,u}$, where
\begin{equation}
(C_f)_{i,u}=-\rho \cdot Sf_{u}\cdot \operatorname{log}(\widetilde{Sf}_{u}),
\label{m1:costsf}
\end{equation}
\begin{equation}
(C_v)_{i,u}=-[N_D\cdot v_u\cdot \operatorname{log} \tilde{v}_i+N_D(1-v_u)\cdot \operatorname{log} (1-\tilde{v}_i)],
\label{m1:costv}
\end{equation}
\begin{equation}
(C_m)_{i,u}=-[n_{c}\cdot m_u^* \cdot \operatorname{log}\tilde{m}_i^*+n_{c}(1-m_u^*)\cdot \operatorname{log}(1-\tilde{m}_i^*)].
\label{m1:costm}
\end{equation}
The assignment problem can be expressed as follows,
\begin{equation}
\textbf{SP}=\underset{\textbf{SP}\in\mathcal{SP}}{\texttt{argmin}}\operatorname{tr} (\textbf{SP}^T(C_v+C_f+C_m))
\label{m1:unbalance}
\end{equation}
This problem can be effectively addressed using the Hungarian algorithm~\cite{Hungarian}, whose complexity can be optimized to $O(n\cdot n_s + {n_s}^2\cdot \operatorname{log}n_s)$ in the unbalanced case as shown in~\cite{Hungarian-complex}. We formally describe the module in Algorithm \ref{alg:1}.
\begin{algorithm}[t!]
\small
\caption{Recovery for the \texttt{s-term} of queries.}
\label{alg:1}
\KwIn{Encrypted database EDB, keyword universe $\Delta_k$, keyword combination universe ${\Delta_c}$, a query list $Q^{\rho}$, Number vector of keyword combinations with the same \texttt{s-term} $\tilde{\textbf{m}}$, volume vector of keywords $\tilde{\textbf{v}}$, \texttt{s-term} query frequency vector of all keywords $\widetilde{\textbf{Sf}}$}
\KwOut{A map from \texttt{s-term} tokens of $Q^{\rho}$ to their underlying keyword SP}

Extract \texttt{s-term} token $SQ^{\rho}\gets Q^{\rho}$\\
Partition $Q^{\rho}$ to $Q_1,\ldots,Q_{n_s}$ and $\Delta_c$ to ${\Delta^1_c},\ldots,{\Delta^n_c}$ according to their \texttt{s-term}.\\
Abstract \texttt{s-term} token universe $\Delta_{\gamma}=[\gamma_1,\ldots,\gamma_{n_s}]$.\\
$Sf,v,m,m^*\gets SQ^\rho$. \\
Compute $C_f,C_v,C_m$.\\
Get the mapping of \texttt{s-term} query to keywords \textbf{SP} by solving the linear assignment problem:\newline $\textbf{SP}=\underset{\textbf{SP}\in\mathcal{SP}}{\texttt{argmin}}\operatorname{tr} (\textbf{SP}^T(C_v+C_f+C_m))$.\\
\Return $\textbf{SP}$
\end{algorithm}
\subsection{Candidate Keyword Conjunction Pruning}

Based on the above observation, we design \textit{CandiPrun}, which leverages the correlation between keywords in conjunctive queries to prune exponentially large candidate keyword conjunctions. The core insight is to prioritize keyword conjunctions with higher query likelihood using \texttt{s-term}-conditioned relative frequencies, thereby reducing the search space while retaining high-probability candidates. 
\textit{CandiPrun} includes two parts, pruning the candidate keyword conjunction and updating the attacker's auxiliary knowledge.


\noindent \textbf{Threshold-Based Pruning with Parameter Tuning. } 
Pruning the candidate set faces a trade-off problem: If the pruning ratio is too high, a large number of queried keyword conjunctions will be removed from the candidate set, leading to extremely low attack accuracy. Conversely, if the pruning ratio is too low, a vast number of useless keyword conjunctions will interfere with query recovery, which not only increases the time and space overhead of the attack but also reduces its accuracy. To achieve an appropriate pruning effect, we set a threshold $\frac{1}{\rho} \times frac$, where $\rho$ is the total number of queries, and $frac \in (0,1]$ is a tunable parameter controlling the strictness of pruning. Let \( f(z^i_j) \) denote the raw query frequency of the conjunction, where the sum of the raw query frequencies of all keyword combinations in \(\Delta_c\) equals 1. Keyword conjunctions with $f(z^i_g) > \frac{1}{\rho} \times frac$ are retained as candidates. This threshold is derived from the observation that low-correlation conjunctions (with $f(z^i_g) \ll \frac{1}{\rho} \times frac$) contribute negligibly to actual query patterns, as validated by the Zipf's distribution in Figure \ref{fig:m2obs1}-\ref{fig:m2obs2}.

We sort all keyword conjunctions by $f(z^i_g)$ in descending order and select the top-$k_i$ candidates for each \texttt{s-term} $w_i$, where $k_i$ is the size of the filtered set. The filtering ratio $\beta_i = \frac{k_i}{\tilde{m_i}}$ measures the pruning efficiency,with $\tilde{m}_i$ denoting the original number of candidate conjunctions for $w_i$. For example, if $\tilde{m}_i= 10^4$, $k_i=200$, and $\beta_i=0.02$,indicating a 98\% reduction in the search space.


\noindent \textbf{Updating Attacker’s Auxiliary Knowledge. }
After pruning, we obtain the filtered candidate set ${\Delta^i_c}^{'}=[z^i_1,\ldots,z^i_{k_i}]$, we then compute the relative query frequency for each keyword conjunction $z^i_g$ containing the \texttt{s-term} $w_i$. Let $f(z^i_j)$ denote the raw query frequency of the conjunction, and $f(w_i)=\sum_{z^i_g\in {\Delta^i_c}^{'}}f(z^i_g)$ denote the total frequency of filtered conjunctions involving the \texttt{s-term} $w_i$. The relative frequency is then normalized as $\text{Pr}(z^i_g|w_i)=\frac{f(z^i_g)}{f(w_i)}$, where $\text{Pr}(z^i_g|w_i)$ represents the conditional probability of querying the keyword conjunction $z_i$ given that the \texttt{s-term} $w_i$ is already known. This normalization is within the same \texttt{s-term} context, reflecting the actual relevance of $z_i$ to $w_i$ in real-world queries.

We further update the attacker's auxiliary knowledge, transforming the frequency vector $\widetilde{\textbf{f}_i}$ and the volume matrix $\widetilde{\textbf{V}_i}$ into pruned versions $\widetilde{\textbf{f}_i}^{'}$ and $\widetilde{\textbf{V}_i}^{'}$. These updated versions focus only on high-probability conjunctions, significantly reducing the computational complexity for the subsequent module. The scaling factor $\beta=[\beta_1,\ldots,\beta_n]$ captures the intensity of pruning in all \texttt{s-term}s, allowing the attacker to balance between the reduction of search space and the retention of information. Since only the top-k items are considered latter, we scale the relative \texttt{s-term} combination pattern of the auxiliary information in \textit{SRecover} by \textbf{$\beta$}. 

This pruning step is critical for the feasibility of the practical attack: by exploiting real-world query correlations \textit{CandiPrun} reduces the exponential $O(n^d)$ search space to a manageable $O(k\cdot n)$, where $k$ is the average number of candidates after pruning, making full query reconstruction feasible even for high-dimensional conjunctive search schemes. 



\subsection{Full Query Reconstruction}
In this module \textit{FullRecover}, we leverage the \texttt{s-term} recovered in \textit{SRecover} and the new candidate keyword conjunctions pruned in \textit{CandiPrun} to further reconstruct the full queries. 

For all queries after the same \texttt{s-term} partition, recall that we have recovered the \texttt{s-term} token $\gamma_u$ with keyword $w_i$ in \textit{SRecover} and a new candidate set ${\Delta^i_c}^{'}$ has been obtained by filtering the keyword conjunctions in \textit{CandiPrun}. In this module, we try to recover all the full queries under each \texttt{s-term}, whose length is denoted as $\rho_u$. We look for the mapping $\textbf{P}_u$ that maximizes the likelihood of observing $\textbf{f}_u$, $\textbf{V}_u$, $\rho_u$ and $N_D$ given the auxiliary information $\widetilde{\textbf{f}_i}^{'}$ and $\widetilde{\textbf{V}_i}^{'}$. Formally, it solves the maximum likelihood problem
\begin{equation}
\textbf{P}_u=\underset{\textbf{P}_u\in\mathcal{P}_u}{\texttt{argmax}} \operatorname{Pr}(\textbf{f}_u,\rho_u, \textbf{V}_u, N_D \mid \widetilde{\textbf{f}_i},\widetilde{\textbf{V}_i},\textbf{P}_u).
\label{m3:argmax}
\end{equation}

We still transform this maximum likelihood problem into minimizing the negative log-likelihood. However, this optimization problem with respect to $\textbf{P}_u$ is not entirely linear, as it includes both linear and quadratic terms. For linear terms, we can use the Hungarian algorithm to solve our optimization problem. For quadratic terms, we can apply the iterative heuristic solution method for quadratic optimization problems proposed in~\cite{usenix22-ihop} to solve. Specifically, we first express the leakage that the attacker can obtain, then we explain how to combine them to fit our attack. 
\begin{algorithm}[htbp]
\small
\caption{Recovery for all the entire queries.}
\label{alg:3}
\KwIn{Encrypted database EDB, keyword universe $\Delta_k$, filtered candidate  universe with $i$-th \texttt{s-term} ${\Delta^i_c}^{'}$, a query list $Q_u$ of \texttt{s-term} token $\gamma_u$, volume matrix $\widetilde{\textbf{V}_i}^{'}$ and query frequency vector $\widetilde{\textbf{f}_i}^{'}$ of keyword combinations in ${\Delta^i_c}^{'}$}
\KwOut{A map from $Q_u$ to $z \in {\Delta^i_c}^{'}$}

$f_u,V_u\gets Q_u$.\\
Get the initial mapping $\textbf{P}_u$ by solving the linear assignment problem:\newline
$\textbf{P}_u=\underset{\textbf{P}_u\in\mathcal{P}_u}{\texttt{argmin}} \sum\limits_{z^i_g\in{\Delta^i_c}^{'}}\sum\limits_{\tau^u_j\in{\Delta^u_{\tau}}}(B^1_{V_u}+B^1_{f_u})_{g,j}\cdot ({\textbf{P}_u})_{g,j}$\\

\For{round from 1 to $n_{iter}$}{
    ${{\Delta^u_{\tau}}}^{\circ} \xleftarrow{\lceil p_{free}\rceil} {\Delta^u_{\tau}}$\\
    ${{\Delta^u_{\tau}}}^{\bullet}=\{\tau^u_j|\tau^u_j\in{\Delta^u_{\tau}},\tau^u_j\notin {\Delta^u_{\tau}}^{\circ}\}$\\
    ${{\Delta^i_c}^{'}}^{\bullet}=\{z^i_g|g=p_u(j), z^i_g\in{\Delta^i_c}^{'},\tau^u_j\in {\Delta^u_{\tau}}^{\bullet}\}$\\
    ${{\Delta^i_c}^{'}}^{\circ}=\{z^i_g|z^i_g\in {\Delta^i_c}^{'},z^i_g \notin {{\Delta^i_c}^{'}}^{\bullet}\}$\\
    $\textbf{P}_u^{\bullet}=\{\tau^u_j\to z^i_{p_u(j)}|\tau^u_j\in {\Delta^u_{\tau}}^{\bullet}\}$\\
    Get $\textbf{P}_u^{\circ}$ by solving the linear assignment problem:\newline $\textbf{P}_u^{\circ}=\underset{\textbf{P}_u^{\circ}\in\mathcal{P}_u^{\circ}}{\texttt{argmin}}\sum\limits_{z^i_g\in{{\Delta^i_c}^{'}}^{\circ}}\sum\limits_{\tau^u_j\in{{\Delta^u_{\tau}}}^{\circ}}(\sum\limits_{\tau^u_{j'}\in{\Delta^u_{\tau}}^{\bullet}}\sum\limits_{z^i_{g'}\in{{\Delta^i_c}^{'}}^{\bullet}}$\newline    $(B^2_{V_u})_{g,g',j,j'}\cdot ({\textbf{P}_u^{\circ}})_{g,j} \cdot ({\textbf{P}_u^{\bullet}})_{g',j'}+(B^1_{V_u}+B^1_{f_u})_{g,j}\cdot ({\textbf{P}_u^{\circ}})_{g,j})$\\
    $\textbf{P}_u \gets \text{combine}({\textbf{P}_u^{\circ}},{\textbf{P}_u^{\bullet}})$\\
}
\Return $\textbf{P}_u$
\BlankLine
\end{algorithm}

\noindent \textbf{Global search pattern leakage.}
Note that the correlations we consider are only between the keywords within the same conjunctive query, while each query is treated independently. As a result, the optimization term related to query frequency remains linear. 

Recall that $\textbf{f}_u$ is the vector of observed entire query token frequencies, $\rho_u$ is the number of queries with \texttt{s-term} token $\gamma_u$, and $\widetilde{\textbf{f}_i}^{'}$ is the vector of auxiliary candidate keyword combination frequencies with \texttt{s-term} $w_i$. We use a Poisson model to compute the attack coefficients. We assume that, when the keyword conjunction $z^i_g$ assigns to the query token $\tau^u_j$, the number of times the user sends token $\tau^u_j$ follows a Poisson distribution with rate $\rho_u\cdot \widetilde{\textbf{f}_i}^{'}_g$. Thus, the log-likelihood cost of assigning $z^i_g\to\tau^u_j$ is $(B_{f_u}^1)_{g,j}=-\operatorname{log Pr(Pois}(\rho_u\cdot \widetilde{\textbf{f}_i}^{'}_g)=\rho_u\cdot {\textbf{f}_u}_j)$. Expanding the expression and ignoring the additive terms
, we get
\begin{equation}
(B^1_{f_u})_{g,j}=-\rho\cdot {\textbf{f}_u}_j \cdot \operatorname{log} \widetilde{\textbf{f}_i}^{'}_g.
\label{m3:fu}
\end{equation}

\noindent \textbf{Global access pattern leakage.}
We express global access pattern leakage in volume form, which is considered specifically the ratio of the number of files matched by two query tokens to the total number of documents. It takes into account both the cost allocation for identical tokens and the joint cost allocation for different query tokens, therefore, the optimization term related to query volume includes both linear and quadratic components. 

Recall that $\textbf{V}_u$ is the matrix of observed volumes of the entire tokens whose \texttt{s-term} token $\gamma_u$ has already assigned to the keyword $w_i$, and $\widetilde{\textbf{V}_i}^{'}$ is the matrix of corresponding auxiliary keyword conjunction volumes. We use a binomial model to get the coefficients $B_{V_u}^1$ and $B_{V_u}^2$. We assume that when keyword conjunction $z^i_g$ assign to query token $\tau^u_j$, the number of documents matching token $\tau^u_j$ follows a binomial distribution with $N_D$ trials and probability given by the auxiliary keyword conjunction volume $({\widetilde{\textbf{V}_i}^{'}})_{g,g}$. Thus, the log-likelihood cost of assigning $z^i_g\to\tau^u_j$ is $(B^1_{V_u})_{g,j}=-\operatorname{log Pr(Bino}(N_D,({\widetilde{\textbf{V}_i}^{'}})_{g,g})=N_D\cdot({\textbf{V}_u})_{j,j})$. Expanding the expression and ignoring the additive terms, we get 
\begin{equation}
\footnotesize
(B^1_{V_u})_{g,j}=-N_D[({\textbf{V}_u})_{j,j} \operatorname{log}({\widetilde{\textbf{V}_i}^{'}})_{g,g}-(1-({\textbf{V}_u})_{j,j})\operatorname{log}(1-{(\widetilde{\textbf{V}_i}^{'}})_{g,g})].
\label{m3:Vu1}
\end{equation}
When $z^i_g\to\tau^u_j$ and $z^i_{g'}\to\tau^u_{j'}$, the number of documents that match both tokens $\tau^u_{j}$ and $\tau^u_{j'}$ follows a Binominal distribution with $N_D$ trials and probability $({\widetilde{\textbf{V}_i}^{'}})_{g,g'}$, we get
\begin{equation}
\footnotesize
(B^2_{V_u})_{g,g',j,j'}=-N_D[({\textbf{V}_u})_{j,j'} \operatorname{log}({\widetilde{\textbf{V}_i}^{'}})_{g,g'}-(1-({\textbf{V}_u})_{j,j'})\operatorname{log}(1-{(\widetilde{\textbf{V}_i}^{'}})_{g,g'})].
\label{m3:Vu2}
\end{equation}

\noindent \textbf{Leakage combinations.}
We consider the scheme with both global search pattern and global access pattern. We combine them by adding their coefficients to fit our attack. The underlying idea is that with log-likelihoods, adding coefficients corresponds to multiplying probabilities. At this point, we can apply the iterative heuristic approach to solve quadratic optimization problems proposed by IHOP~\cite{usenix22-ihop} to address our modeled problem. We formally describe this module in Algorithm \ref{alg:3}.

\section{Performance Evaluation}
\label{sec:evaluation}

\subsection{Experimental Setup}
We use Python 3.9.13 to implement all experiments and run them on a laptop using an Intel(R) Core(TM) i5-8265U CPU@1.60GHz with 8GB RAM.

\noindent \textbf{Datasets. }
We conduct our experiments using two widely-used datasets: the Enron email corpus and the Lucene java-user mailing list. The Enron\footnote{\href{https://www.cs.cmu.edu/~./enron}{https://www.cs.cmu.edu/~./enron}} dataset, collected between 2000 and 2002, comprises 30,109 emails. The Lucene\footnote{\href{https://lucene.apache.org/}{https://lucene.apache.org/}} dataset consists of 66,491 emails from the java-user mailing list. For the Enron dataset, we select the 50, 100, 200, and 300 most frequent words, excluding those affected by stemming, as keyword universe. These keywords and their associated identifiers are used to construct the encrypted database. Since it is difficult to obtain the conjunctive keyword query frequency on Google Trend, we adopt a different approach for the Lucene dataset. Specifically, to expand the keyword set to a total of 300, we first select from the top 300 most frequent words in Enron that are also present in Lucene. For the remaining slots, we randomly sample additional terms from Lucene's vocabulary while excluding any already selected terms.

\noindent \textbf{CSSE schemes. }
We mainly focus on state-of-the-art CSSE schemes~\cite{crypto13-oxt, ccs18-hxt, ccs24-doris}, which build upon the OXT framework. 
Note that many schemes~\cite{vldb19-vbtree,icde17-ibtree,tdsc24-guo}, exhibit broader leakage profiles compared to OXT framework, rendering them also vulnerable to our attack.

\noindent \textbf{Frequency information. }
We collect 260 weeks of data from Google Trends\footnote{\href{https://trends.google.com/trends/}{https://trends.google.com/trends/}}, spanning January 2019 to December 2023. It consists of two components: the query frequency for each individual keyword and the conjunctive query frequency for each 2-dimensional keyword conjunction. Due to the exponential growth of keyword conjunctions as the dimension $d$ increases, extracting conjunctive query frequencies for all keyword conjunctions becomes computationally infeasible. Therefore, we limit our data collection to conjunctive query frequencies for only 2-dimensional keyword conjunctions and employ frequency approximation methods to estimate conjunctive query frequencies for higher-dimensional conjunctions ($d>2$). The frequency approximation method used is detailed in Appendix \ref{app:freq_approx}. 
To simulate user queries, we calculate the sum of query frequencies over weeks 211 to 260 and normalize the frequency of each keyword conjunction by dividing the frequency sum of all keyword conjunctions. For the attacker’s auxiliary knowledge, we derive the corresponding summed frequencies over weeks $211-T$ to $260-T$, where $T$ represents the temporal offset between the attacker's knowledge and the user’s observation.

\noindent \textbf{Attacker's knowledge.} The same as~\cite{usenix24-jigsaw}, we assume the attacker has knowledge of a similar dataset and partition the document set into two disjoint subsets with equal size. One subset is selected as the client's encrypted database, while the other serves as the attacker's auxiliary knowledge, representing a similar dataset. In contrast to stronger known-data assumptions adopted in~\cite{ccs21-leap,ccs15-count,ndss20-revisiting}.
Each experiment is performed over 10 independent runs to ensure the reliability of the results.

\noindent \textbf{Metrics. }
To precisely describe the LAAs in conjunctive keyword query scenarios, we extend the \textit{accuracy} metric used in single keyword query scenarios. Specifically, we propose four evaluation metrics:  \textit{\texttt{s-term} recovery accuracy (s-acc)}, \textit{full query recovery accuracy (f-acc)}, \textit{loose query recovery accuracy (l-acc)} and \textit{cumulative accuracy distribution (CAD)}. For a list of attacked queries, these metrics are defined and computed as follows.
\begin{itemize}
\item[$\bullet$] \textit{\texttt{s-term} recovery accuracy (s-acc)}: The proportion of queries for whose \texttt{s-term} is correctly recovered. \\ $s\mbox{-}acc=\frac{\#\;queries\; with\;s\mbox{-}term\;recovered\;correctly}{\#\;total\;queries}$
\item[$\bullet$] \textit{Full query recovery accuracy (f-acc)}: The proportion of queries in which all keywords involved in the query are fully recovered. $f\mbox{-}acc=\frac{\#\: queries\; with\;all\;keywords\;recovered\;correctly}{\# \;total\;queries}$
\item[$\bullet$] \textit{Loose query recovery accuracy (l-acc)}: The proportion of total queried keywords that are correctly recovered.\\ $l\mbox{-}acc=\frac{\#\;keywords\; which\;recovered\;correctly}{\#\;total\;keywords}$
\item[$\bullet$] \textit{Cumulative accuracy distribution (CAD)}: Inspired by cumulative probability distribution, this metric evaluates the proportion of queries for which at least $x$ keywords are recovered. \\${CAD}_x=\frac{\#\;queries\; with\;at\;least\;x\;keywords\;recovered\;correctly}{\#\;total\;queries}$
\end{itemize}

\subsection{Results of {\sffamily S-Leak} Attack} 
\label{sec:exp_similardata}
\begin{figure*}[htbp]
        \centering 
        \begin{subfigure}[b]{0.31\textwidth}
            \label{fig:enron_rho_d2_separate}
            \includegraphics[width=\textwidth]{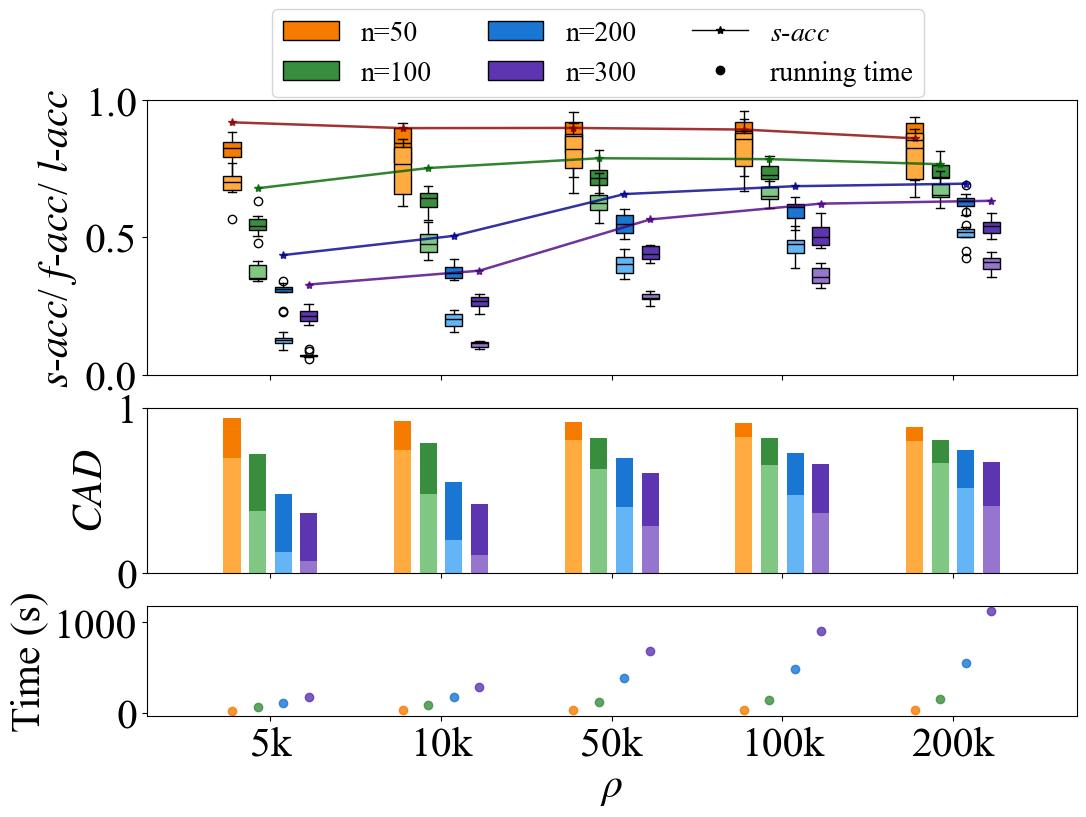}
            \caption{\normalfont Separate, Enron $d=2$}
        \end{subfigure}
        \hspace{10pt}
        \begin{subfigure}[b]{0.31\textwidth}
            \label{fig:lucene_rho_d2_separate} 
            \includegraphics[width=\textwidth]{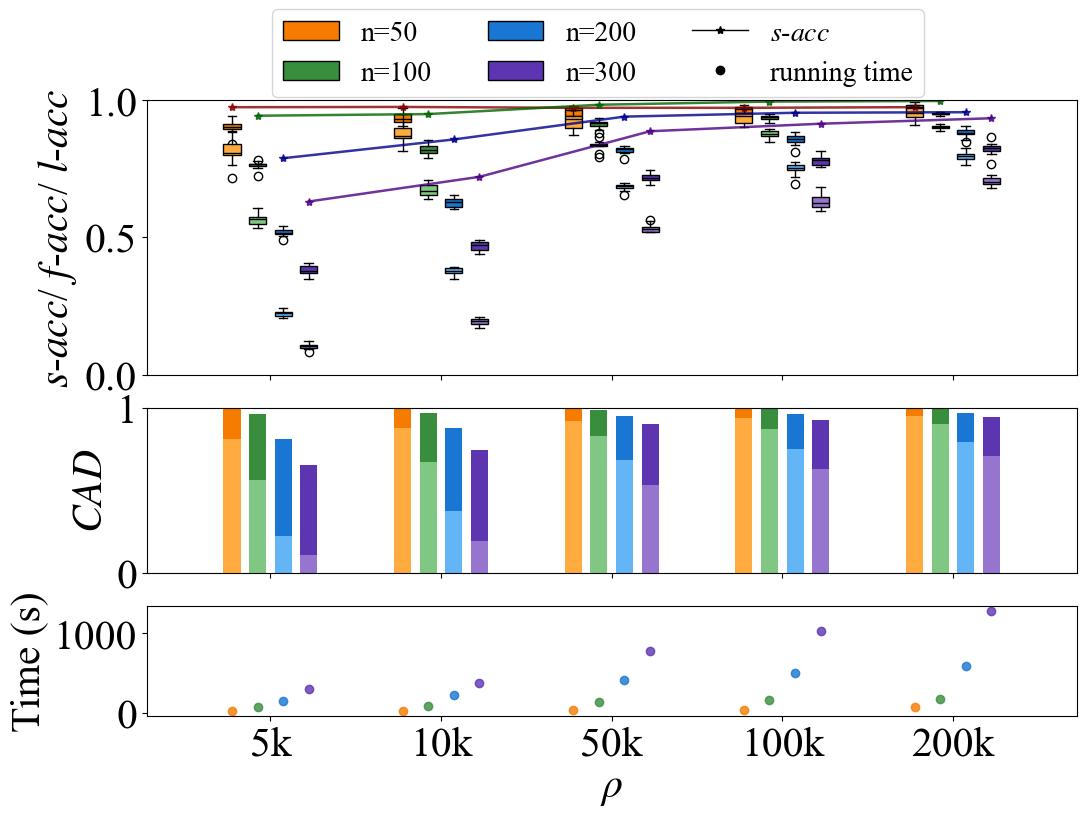}
            \caption{\normalfont Separate, Lucene $d=2$}
        \end{subfigure}
        \hspace{10pt}
        \begin{subfigure}[b]{0.312\textwidth}
            \label{fig:enron_rho_d2_hybrid}
            \includegraphics[width=\textwidth]{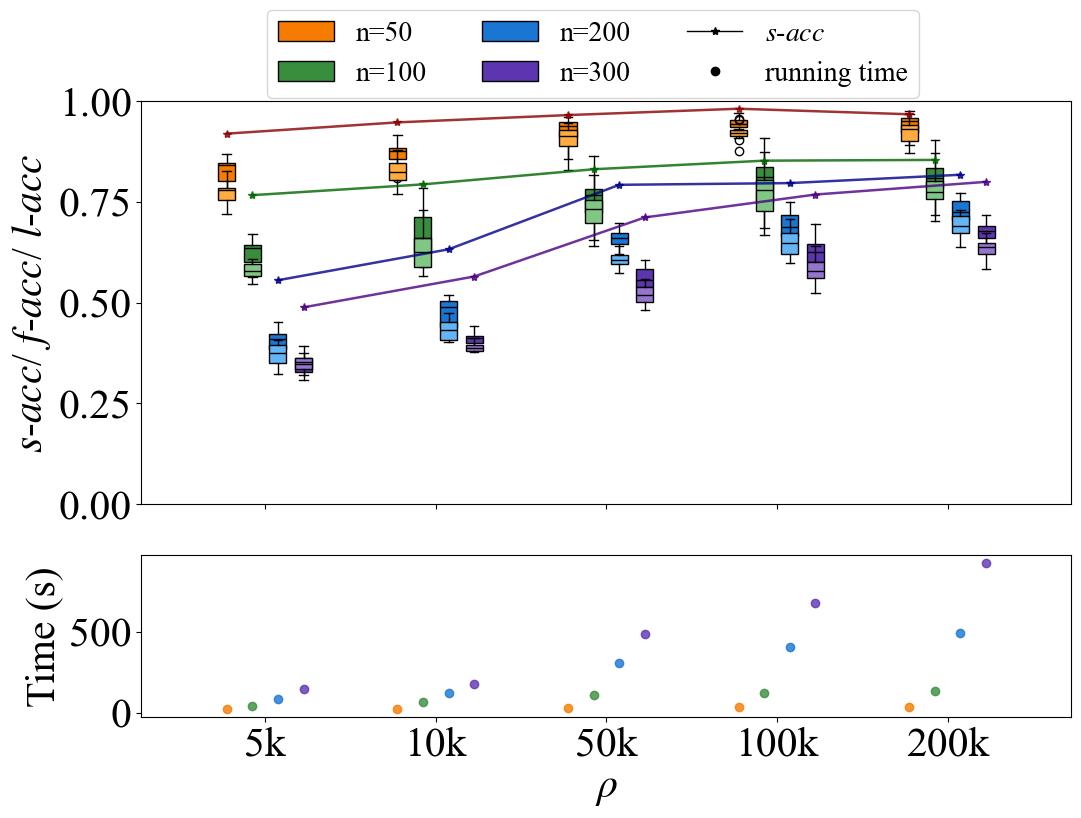}
            \caption{\normalfont Hybrid, Enron $d=2$}
        \end{subfigure}
        \begin{subfigure}[b]{0.31\textwidth}
            \label{fig:enron_rho_d3_separate}
            \includegraphics[width=\textwidth]{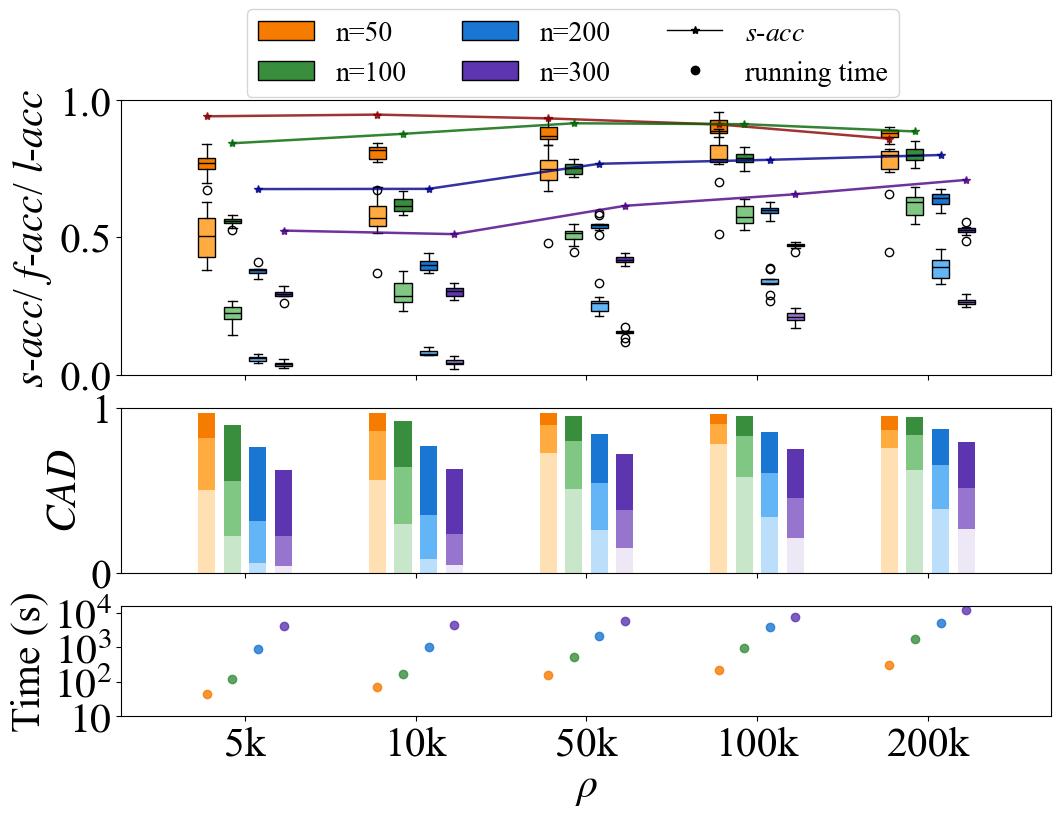}
        \caption{\normalfont Separate, Enron $d=3$}
        \end{subfigure}
        \hspace{10pt}
        \begin{subfigure}[b]{0.31\textwidth} 
            \label{fig:lucene_rho_d3_separate} 
            \includegraphics[width=\textwidth]{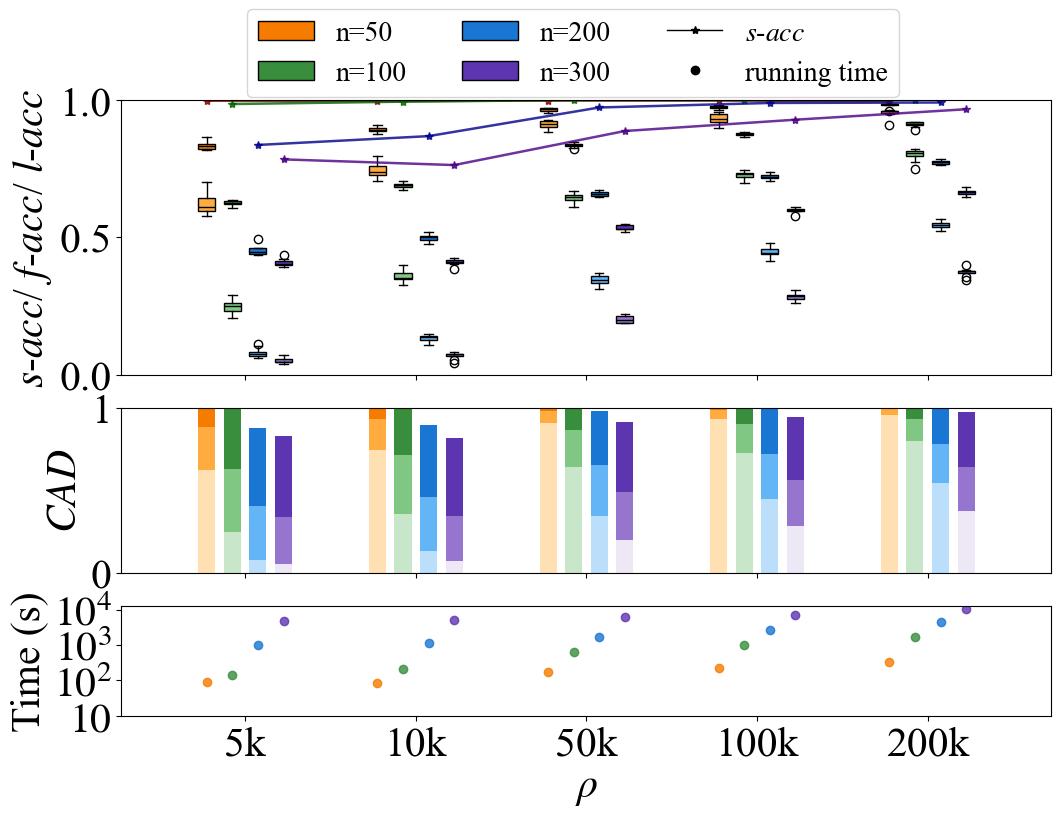}
        \caption{\normalfont Separate, Lucene $d=3$}
        \end{subfigure}
        \hspace{10pt}
        \begin{subfigure}[b]{0.312\textwidth}
            \label{fig:enron_rho_d3_hybrid}
            \includegraphics[width=\textwidth]{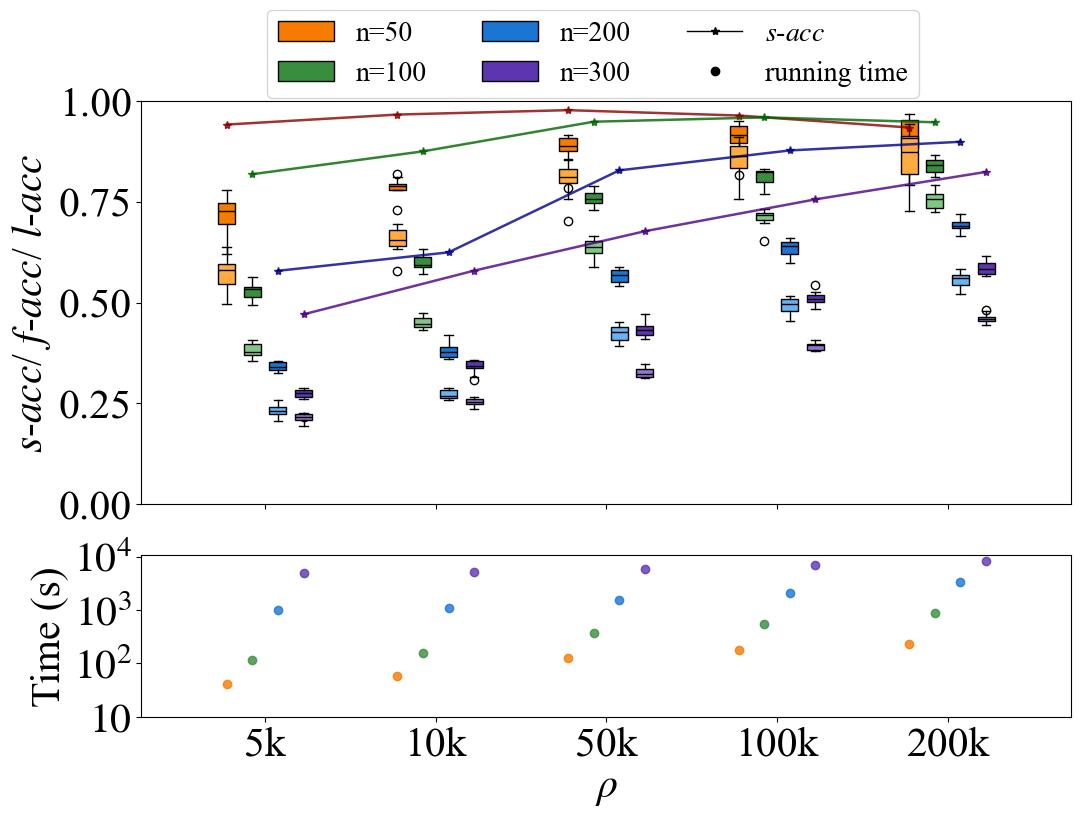}
            \caption{\normalfont Hybrid, Enron $d=3$}
        \end{subfigure}
        
        \caption{Performance of {\sffamily S-Leak}. Each column comprises two vertically aligned boxplots: the upper boxplot corresponds to the $l\mbox{-}acc$, while the lower one represents the $f\mbox{-}acc$. The stacked bar chart below illustrates the $CAD$ of the attack, the color gradient from dark to light corresponds to the recovery of at least 1 keyword, at least 2 and 3 keywords, respectively.} 
        \label{fig:res_rho} 
\end{figure*}
In this subsection, we investigate the impact of the query volume $\rho$ on the attack performance with 2-dimensional ($d=2$) and 3-dimensional ($d=3$) conjunctive queries. The empirical validation leverages leakage patterns identified in Section~\ref{sec:pre_leakage} to demonstrate {\sffamily S-Leak}'s overall effectiveness under two distinct query settings: separate query (queries with fixed dimensions) and hybrid query (queries with varying dimensions).

Specifically, we select 50, 100, 200, 300 keywords as the keyword universe. And for both settings, we measure the attack accuracy using three metrics: \texttt{s-term} recovery accuracy ($s\mbox{-}acc$), full query recovery accuracy ($f\mbox{-}acc$), loose query recovery accuracy ($l\mbox{-}acc$). We compute the cumulative accuracy distribution ($CAD$) only in the separate query setting, as varying dimensions in hybrid query setting makes $CAD$ harder to compare. We also analyze the running time of the attack to evaluate the efficiency of our attack. 

To balance accuracy and computational cost, we utilize $frac=~0.6$ (the impact of $frac$ on attack performance is further explored in Section \ref{sec:exp_abl}) to pruning candidate conjunctions and set parameters in the third module with $n_{iter}$=1000 and $p_{free}$=0.25 (selected based on their demonstrated optimal performance in the experiments of~\cite{usenix22-ihop}.). We set $T=0$ in subsequent experiments, and if we use the past query frequency data, the corresponding accuracy will be slightly lower as discussed in section \ref{sec:exp_dur}.

\noindent \textbf{Result of separate query setting.} 
The results of the separate query setting are shown in Figure~\ref{fig:res_rho}. We observe that as $\rho$ increases, the accuracy also improves. The larger $n$ requires the larger $\rho$ to achieve an attack accuracy comparable to that of the smaller $n$. 
Under $\rho=100,000$ and $n=200$, the separate query setting yields $f\mbox{-}acc=0.4681$ and $l\mbox{-}acc=0.5970$ for $d=2$ on Enron dataset, while the $CAD$ reaches [0.7258, 0.4681], indicating that 72.58\% of queries have at least one keyword recovered and 46.81\% are fully recovered. 
Higher-dimensional queries ($d=3$) exhibit an accuracy decrease in $f\mbox{-}acc$ (e.g., 0.3354 for Enron), while $l\mbox{-}acc$ remains comparable to $d=2$ results (e.g., 0.5976 for Enron). The $CAD$ values [0.8513, 0.6062, 0.3354] observed on Enron dataset reflect the inherent complexity of reconstructing multi-keyword conjunctions. 

Furthermore, we observed the attack achieves higher $s\mbox{-}acc$ for higher-dimensional. This phenomenon can be attributed to two key factors: (1) Combinatorial explosion in high-dimensional keyword conjunctions expands the candidate space. With limited query volume $\rho$, the covered subset centers on high-frequency \texttt{s-term}s, which are more recoverable due to concentrated leakage patterns. (2) Higher-dimensional queries amplify the distinctiveness of \texttt{s-term} leakage patterns, which improve \texttt{s-term} recovery accuracy. 

Under identical experimental conditions, our attack achieves superior performance on Lucene dataset, with $CAD$ reaching [0.9626, 0.7475] for $d=2$ and [0.9926, 0.7195, 0.4462] for $d=3$. This enhanced effectiveness stems from Lucene's larger document corpus, which exhibits stronger volume leakage patterns compared to Enron. 
Progressive recovery of underlying query keywords more effectively reflects real-world scenarios, where recovery even partial keywords of conjunctive queries is sufficient to reveal substantial information. The results demonstrate high accuracy in recovering a small subset of underlying keywords, along with a non-negligible accuracy in recovering all keywords, and do not need any known queries. This underscores the effectiveness of our attack and its potential significance in practical scenarios. 

Regarding running time, we observe that attacks under $d=3$ remain feasible within a reasonable time cost, with only a modest increase in time overhead compared to $d=2$. This is largely due to the design of \textit{CandiPrun}. For the Enron dataset, under $n=200$, $\rho=100,000$ and $frac=0.6$, the fraction parameter $\bar{\beta}=0.677$ for $d=2$, while $\bar{\beta}=0.042$ for $d=3$ is optimized to reduce computational complexity and time overhead in the \textit{FullRecover}. This ensures that even with 4,455,100 possible keyword conjunctions (when $d=3$ and $n=300$), the attack remains practical. 

\noindent \textbf{Result of hybrid query setting.} 
Real-world search systems often process queries with varying dimensions. To model this, we evaluate our attack under the hybrid query setting, where the query dimensions follow a uniform distribution. The result of Enron is presented in Figure~\ref{fig:res_rho}, the results of Lucene are shown in Appendix~\ref{app:add_result} Figure~\ref{fig:res_rho_hybrid_lucene}. The overall trend of attack accuracy under the hybrid setting is consistent with that of the separate query setting. However, compared to the separate query setting, the hybrid setting achieves higher accuracy and incurs a lower time overhead. Specifically, under $\rho=100,000$ and $n=200$, the hybrid query setting yields $f\mbox{-}acc=0.6841$ and $l\mbox{-}acc=0.6870$ for $d=2$ on Enron dataset, and for $d=3$ they are 0.4957 and 0.6396, which is significantly higher than results under the separate query setting ($f\mbox{-}acc=0.4681$ and $l\mbox{-}acc=0.5970$ for $d=2$, 0.3354 and 0.5976 for $d=3$). This is because, given the same number of queries, the hybrid setting involves more queries with lower dimensions, which makes it easier to recover the underlying keywords. 

\subsection{Evaluation on Effectiveness of Modules in {\sffamily S-Leak}}
\label{sec:exp_abl}
In this subsection, we evaluate the effect of the first two modules in {\sffamily S-Leak} on attack accuracy and running time to further demonstrate the effectiveness of our design. 

\begin{figure}[htbp]
    \centering 
    \includegraphics[width=0.35\textwidth]{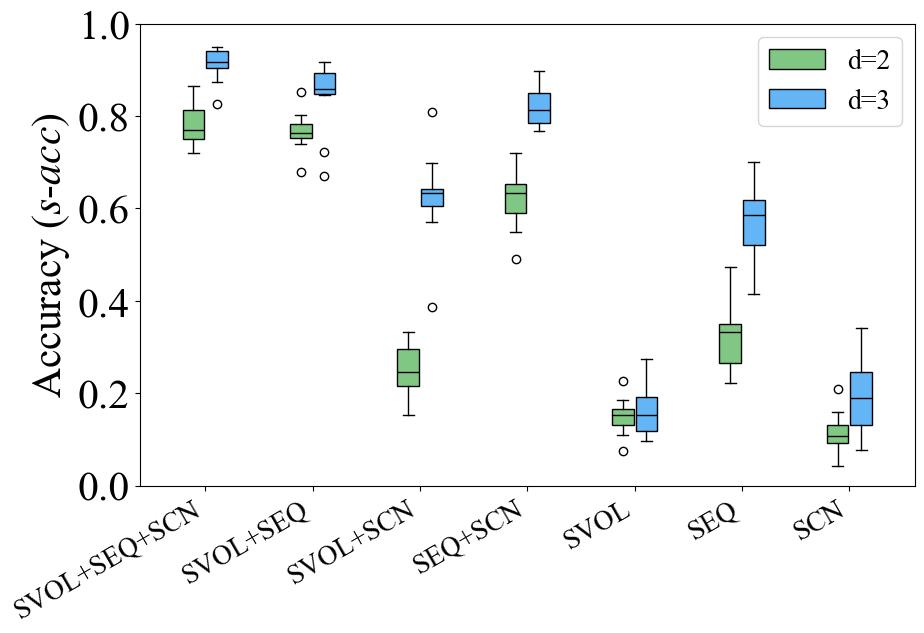} 
    \caption{Effect of different combinations of three \texttt{s-term} leakage patterns on the recovery of \texttt{s-term}s of {\sffamily S-Leak} against CSSE on Enron dataset.} 
    \label{fig:res_m1cost_enron} 
\end{figure}

\noindent \textbf{Effect of three \texttt{s-term} leakage patterns in \textit{SRecover}. }
We first conduct an experiment to investigate the effect of three \texttt{s-term} leakage patterns on the recovery of \texttt{s-term}s. Specifically, we compute the $s$-$acc$ of using individual and combined leakage patterns for \textit{SRecover} of {\sffamily S-Leak} under conditions $n = 100$ and $\rho=100,000$. The results of the Enron dataset, shown in Figure~\ref{fig:res_m1cost_enron}, indicate that using a single pattern for the attack achieves limited effectiveness, while combining multiple leakage patterns significantly improves attack performance. Our experimental results demonstrate that the combined utilization of all three \texttt{s-term} leakage patterns achieves optimal \texttt{s-term} recovery accuracy. The multi-pattern fusion strategy in \textit{SRecover} is pivotal to overcoming the ambiguity of \texttt{s-term} identification. In other experiments, we default use all three leakage patterns jointly for the recovery of \texttt{s-term}. 

\begin{figure}[htbp]
        \centering 
        \begin{subfigure}[b]{0.23\textwidth}
            \label{fig:enron_beta_d2} 
            \includegraphics[width=\textwidth]{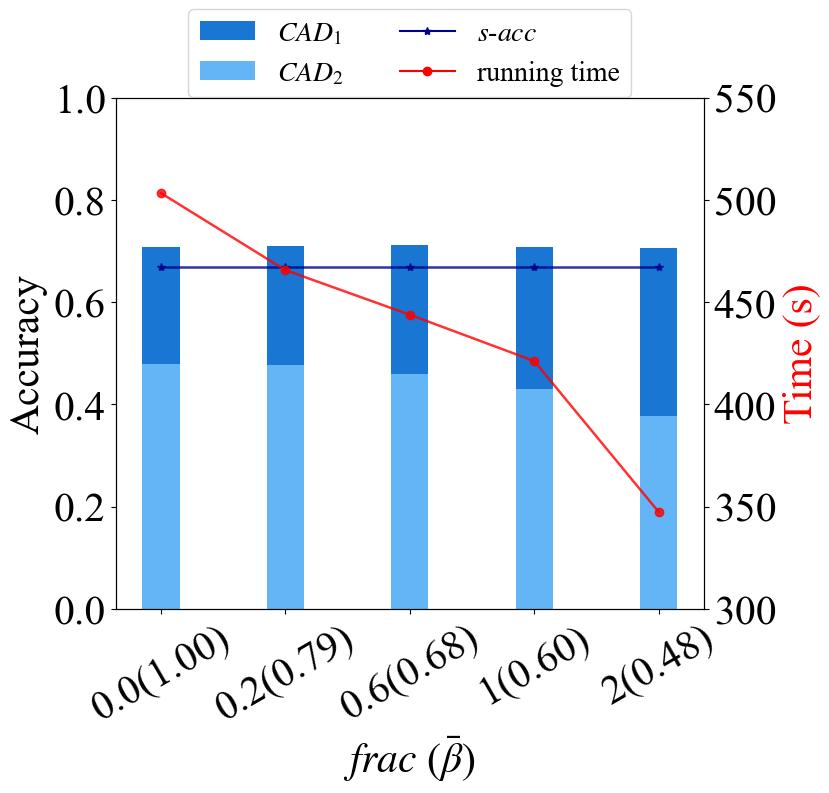}
            \caption{\normalfont $n=200, d=2$}
        \end{subfigure}
        \begin{subfigure}[b]{0.23\textwidth}
            \label{fig:enron_beta_d3} 
            \includegraphics[width=\textwidth]{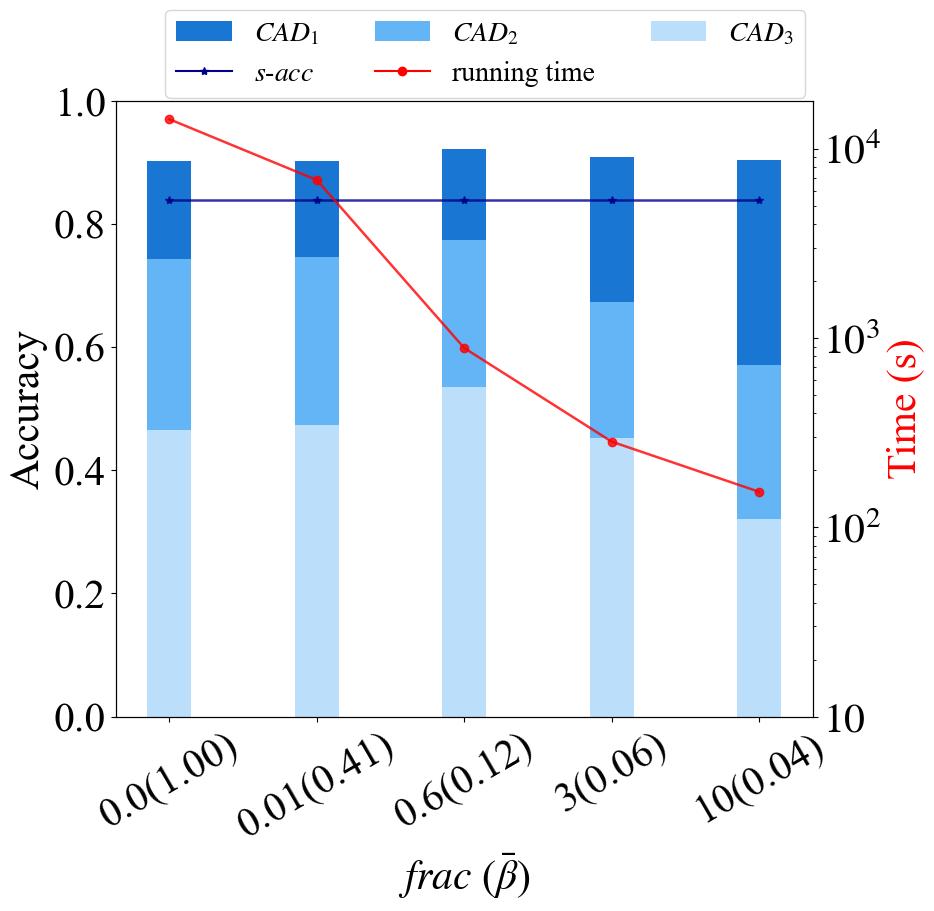} 
            \caption{\normalfont $n=100, d=3$}
        \end{subfigure}
        \caption{Effect of $frac$ ($\bar{\beta}$) in the performance of {\sffamily S-Leak} against CSSE on Enron dataset.} 
        \label{fig:res_beta} 
\end{figure}

\noindent \textbf{Effect of pruning ratio in \textit{CandiPrun}. }
We perform the experiment to analyze how the configuration of the parameter $frac$ (and $\bar{\beta}$) in \textit{CandiPrun} influences both the accuracy ($CAD$) and the running time. To eliminate potential interference from the $frac$-dependent \texttt{s-term} combination pattern optimization process, we restrict \textit{SRecover} in this experiment by using only \texttt{s-term} volume pattern and \texttt{s-term} search pattern. We set $\rho = 100,000$, and evaluate the attack performance under different $frac$ for $n = 200, d = 2$ and $n = 100, d = 3$ on Enron dataset. The results are presented in Figures \ref{fig:res_beta}. We observe that increasing $frac$ within a reasonable range (corresponding to decreased $\bar{\beta}$) effectively reduces the time overhead while maintaining minimal degradation in effectiveness. 

In the case of $d = 3$, the advantage of candidate conjunction pruning becomes even more apparent. When $frac=0$, considering all possible conjunctions, the time cost of the attack increases substantially (reaching 4 hours, which is 15 times that of $frac=0.6$), and the excessive number of candidate keywords negatively affects the accuracy. The pruning mechanism in \textit{CandiPrun} reduces the candidate space by leveraging keyword co-occurrence correlations. For example, with $frac=0.6$, the number of candidate conjunctions decreases from $C(n,3)=1,313,400$ to $\bar{\beta}\cdot C(n,3)=157,608$ for $n=200$, allowing for the feasible recovery of queries of high dimensions. The results indicate that the design of \textit{CandiPrun} greatly reduces the attack overhead, improves the accuracy, and makes the attack still feasible on moderate keyword universe and conjunctive queries with higher dimensions. In other experiments, we set $frac=0.6$, as this value achieves an optimal trade-off in efficiency and effectiveness. 

\subsection{Durability}
\label{sec:exp_dur}
To evaluate the temporal resilience of {\sffamily S-Leak}, we investigate how auxiliary information offset $T$ impacts the attack when using `outdated' query frequency data. This addresses a critical real-world constraint: attackers often lack real-time auxiliary information due to data collection barriers or privacy-preserving countermeasures. By quantifying performance degradation over time, we establish the attack's operational viability in practical scenarios. 
We fix the number of training queries $n=100$ and the total query volume $\rho=100,000$ across all trials, and evaluate recovery accuracy for conjunctive queries with $d=2$ and $d=3$. 

\begin{figure}[htbp]
        \begin{subfigure}[b]{0.234\textwidth}
            \label{fig:freqshift_d2} 
            \includegraphics[width=\textwidth]{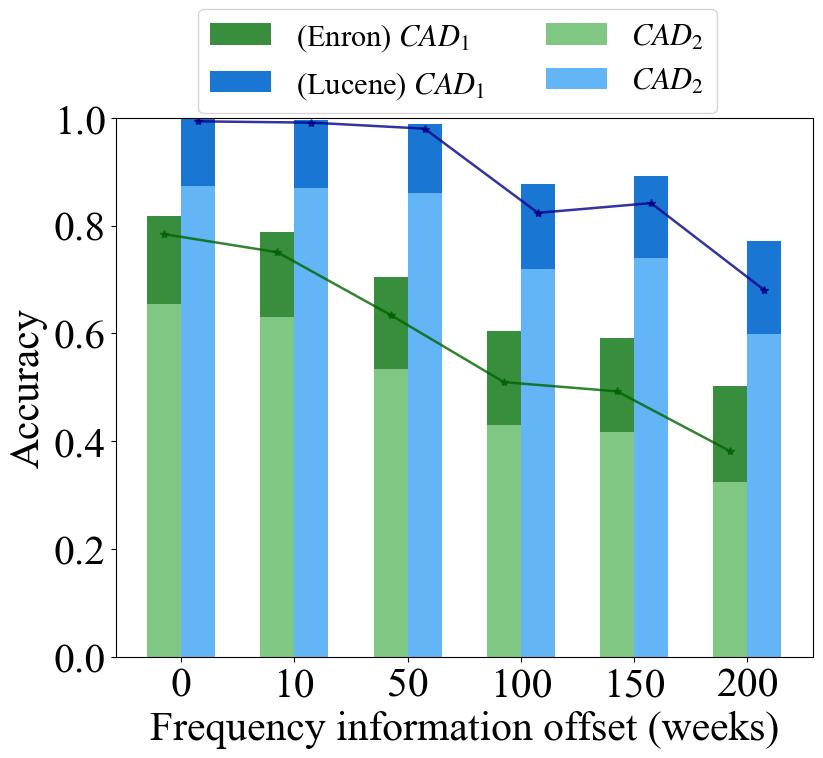}
            \caption{\normalfont $d=2$}
        \end{subfigure}
        \begin{subfigure}[b]{0.234\textwidth}
            \label{fig:freqshift_d3} 
            \includegraphics[width=\textwidth]{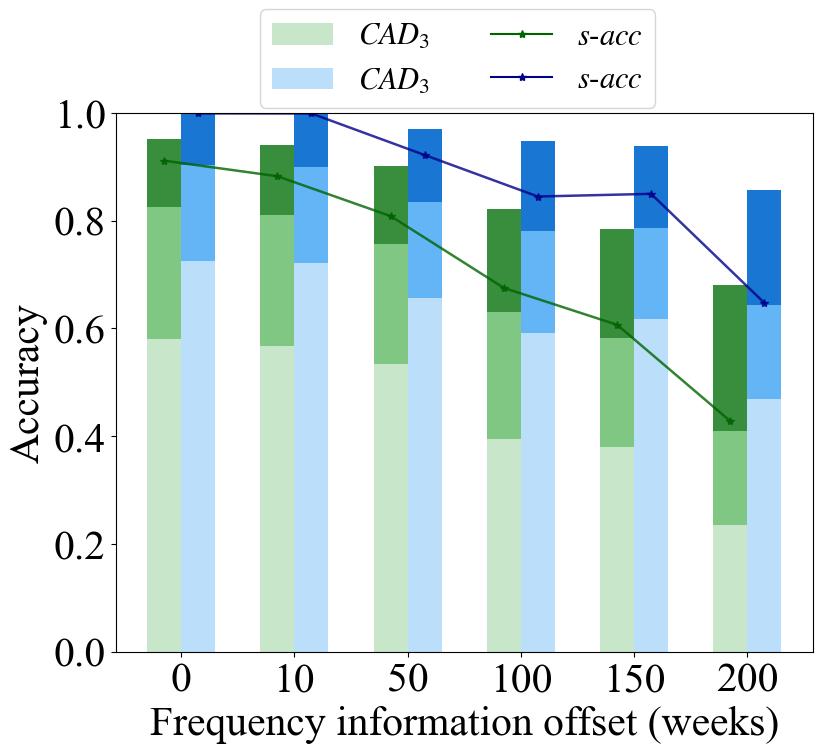}
            \caption{\normalfont $d=3$}
        \end{subfigure}
        \caption{Effect of offset $T$ in the performance of {\sffamily S-Leak} against CSSE on Enron dataset.} 
        \label{fig:res_freqshift} 
\end{figure}

As shown in Figure~\ref{fig:res_freqshift}, the results demonstrate that {\sffamily S-Leak} exhibits temporal resilience even when relying on outdated auxiliary frequency information. Across both Enron and Lucene datasets, the results reveal that increasing the offset $T$ (i.e., using more outdated auxiliary frequency information) leads to a consistent decrease in attack accuracy. The observed results indicating that the freshness of auxiliary data impacts attack performance, however the use of stale data does not render our attack ineffective, especially for recovering at least one keyword ($CAD_1$), with accuracy typically above 50\% even when the offset reaches 200 weeks. And it has enhanced effectiveness on Lucene, where $CAD_1$ maintains over 85\% accuracy for $d=3$ and over 77\% for $d=2$, underscoring the persistent threat posed by frequency leakage in practical settings. These findings suggest that even stale auxiliary data—spanning up to four years—can significantly compromise query privacy, highlighting the need for defenses that mitigate long-term frequency leakage.


\subsection{Against Defenses}
\label{sec:exp_countermeasures}

While several privacy-preserving SSE schemes aim to mitigate leakage-abuse attacks, existing defenses lack systematic evaluation in conjunctive query scenarios. To address this gap, we evaluate our attack against two typical SSE defenses: the padding mechanism in SEAL~\cite{usenix20-seal} and the obfuscation technique in CLRZ~\cite{infocom18-clrz}. Our evaluation considers two adversarial knowledge models: one where the attacker is aware of the client’s deployed defenses (realistic for sophisticated attackers) and a baseline where defense knowledge is absent. We adapt our attack framework to explicitly target padding and obfuscation strategies, with detailed adaptations provided in Appendix~\ref{app:adaptations}. This approach ensures a comprehensive assessment of defense effectiveness in the context of conjunctive queries, where prior work has left critical security gaps unaddressed.

We analyze the performance of our attack against the aforementioned defenses on Enron dataset under a separate query setting. For all defense experiments, we utilize all three \texttt{s-term} leakage patterns to optimize the recovery of \texttt{s-term} tokens, set the candidate filter parameter \( \text{frac} = 0.6 \), and configure the third module with \( n_{\text{iter}} = 1000 \) and \( p_{\text{free}} = 0.25 \). These parameter settings are chosen because they demonstrated the best attack performance in our prior performance evaluation. We further set \( n = 100 \), \( \rho = 100,000 \), and \( T = 0 \) for the defense experiment evaluation.

\begin{figure}[htbp]
        \begin{subfigure}[b]{0.22\textwidth}
            \label{fig:clrz_d2} 
            \includegraphics[width=\textwidth]{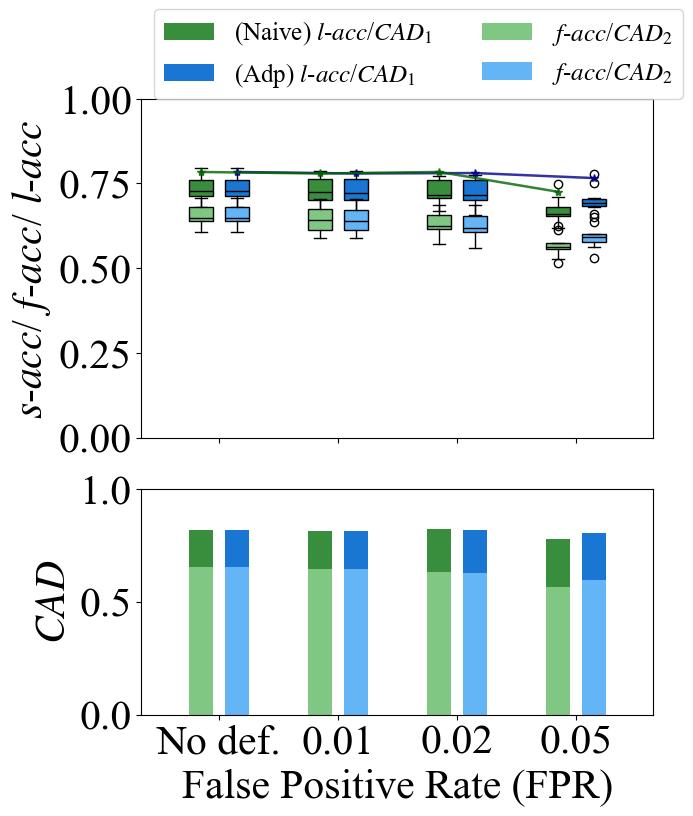}
            \caption{\normalfont $d=2$}
        \end{subfigure}
        \hspace{10pt}
        \begin{subfigure}[b]{0.21\textwidth}
            \label{fig:clrz_d3} 
            \includegraphics[width=\textwidth]{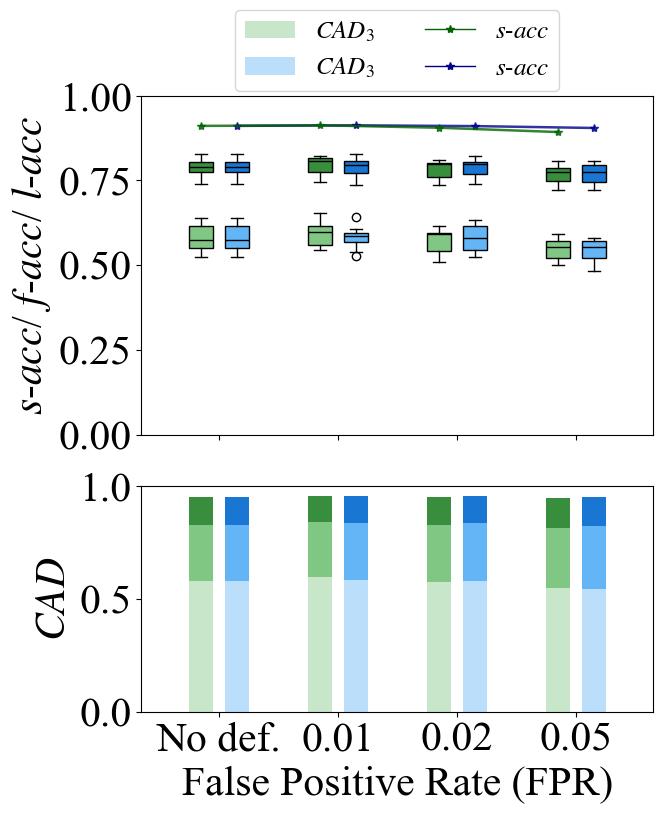}
            \caption{\normalfont $d=3$}
        \end{subfigure}
        \caption{Performance of {\sffamily S-Leak} against the obfuscation in CLRZ~\cite{infocom18-clrz} on Enron dataset.} 
        \label{fig:res_clrz} 
\end{figure}

\noindent \textbf{Against the obfuscation in CLRZ. }
We present experimental results for attacks against the obfuscation method in CLRZ~\cite{infocom18-clrz}. This defense works by associating a keyword with documents that do not contain it with a false positive rate (FPR) and by omitting the index entries for documents that do contain the keyword with a true positive rate (TPR). The obfuscation approach does not use padding, therefore it leaves storage costs unchanged, however, communication costs rise substantially due to the retrieval of many unrelated documents. In Figure \ref{fig:res_clrz}, we set $\text{TPR}=0.999$ and $\text{FPR}\in\{0.01, 0.02, 0.05\}$. This figure shows that under CLRZ obfuscation, the proposed attack exhibits only marginal performance degradation as the false positive rate (FPR) increases, with accuracy reductions consistently remaining below 10\%.

\begin{figure}[htbp]
        \begin{subfigure}[b]{0.22\textwidth}
            \label{fig:seal_d2} 
            \includegraphics[width=\textwidth]{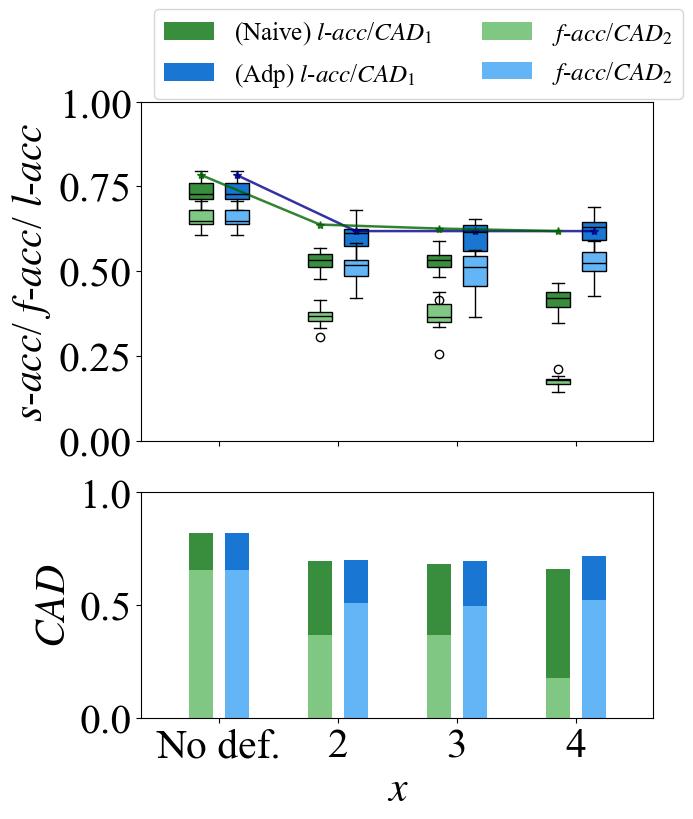}
            \caption{\normalfont $d=2$}
        \end{subfigure}
        \hspace{10pt}
        \begin{subfigure}[b]{0.21\textwidth}
            \label{fig:seal_d3} 
            \includegraphics[width=\textwidth]{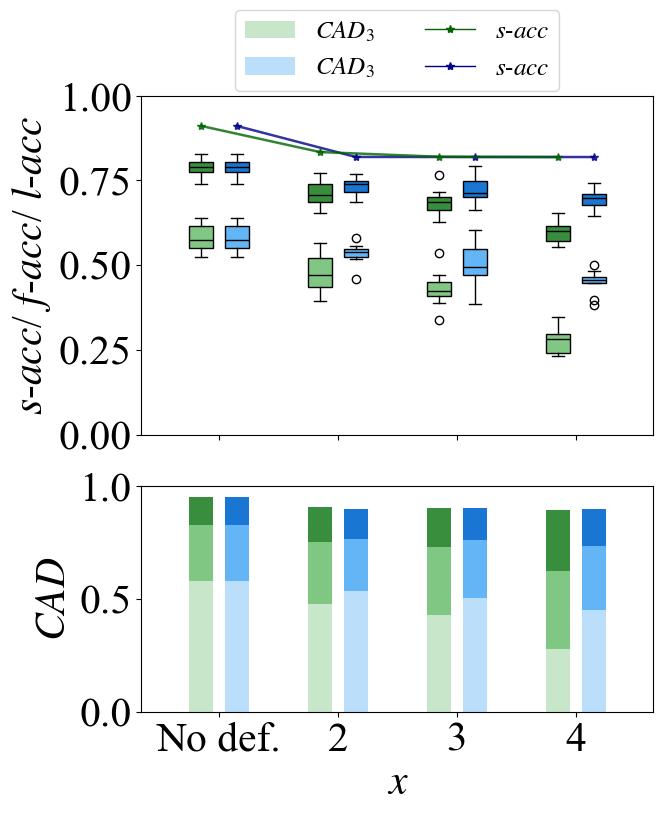}
            \caption{\normalfont $d=3$}
        \end{subfigure}
        \caption{Performance of {\sffamily S-Leak} against the padding in SEAL~\cite{usenix20-seal} on Enron dataset.} 
        \label{fig:res_seal} 
\end{figure}
\noindent \textbf{Against the padding in SEAL}
The SEAL defense mechanism, introduced by Demertzis et al.~\cite{usenix20-seal}, has two adjustable parameters: $\alpha$ and $x$. Under this mechanism, the database is structured into $2^{\alpha}$ ORAM blocks, effectively masking which specific document is accessed within each block during query operations. SEAL further enhances privacy by padding the volume of each query to the closest power of $x$. We vary the padding parameter $x$ between 2, 3 and 4. The experiment results are presented in Figure \ref{fig:res_seal}, which shows that the attack has moderate accuracy reductions ranging from 5\% to 22\%, but maintains substantial accuracy overall.

The limited efficacy of these defenses against our attack is primarily due to they exclusive focus on independent perturbing access patterns and overlook the compounded leakage risks from keyword co-occurrence patterns. Our attack leverages additional leakage patterns, including search patterns and the \texttt{s-term} combination pattern, which collectively improve its resilience to such defenses.

\section{Discussion}
\label{sec:discussion}
In this section, we discuss how to extend our attack to dynamic setting and potential countermeasures. 

\noindent \textbf{Extend to dynamic setting. }
Recent work~\cite{ccs23-dynamic} demonstrates how to break forward and backward privacy by exploiting search pattern and volume information in dynamic settings. Although they focus on single-keyword queries, the same principles can be applied to conjunctive queries by leveraging both \texttt{s-term} and global patterns, to reconstruct the bijection between query tokens and underlying keywords. This reconstruction can then be followed by {\sffamily S-Leak} to complete the attack. It is worth noting that in~\cite{ndss21-odxt}, the definition of \texttt{s-term} changes from the keyword with the least document frequency to the keyword with the least update frequency. In this case, when construct \texttt{s-term} query frequency using auxiliary information, the attacker would need access to certain historical update information from the auxiliary dataset.

\noindent \textbf{Countermeasures. }
As discussed in Section \ref{sec:exp_countermeasures}, existing defenses lack a systematic consideration of multi-keyword conjunctive queries. 
For \texttt{s-term} leakage suppression, the {\ttfamily TSet} structure avoids direct \texttt{s-term} access pattern leakage through its unique design, it still exposes \texttt{s-term} volume patterns, posing non-negligible security risks under our attack evaluation. Mitigation of \texttt{s-term} leakage patterns can be approached through two primary directions: (1) Obfuscate the \texttt{s-term} equality pattern: Introduce ambiguity by assigning multiple token to the same \texttt{s-term} and splitting combination patterns into unlinkable subsets. 
(2) Obfuscate the \texttt{s-term} access/volume patterns: 
Introduce randomness to break the deterministic link between keywords and their leakage patterns. This includes perturbing volume data or randomizing access patterns, making it harder for attackers to exploit statistical consistencies. 

\section{Related Work}
\noindent \textbf{Conjunctive Searchable Symmetric Encryption (CSSE). }
CSSE enables secure multi-keyword document retrieval, with early methods suffering from inefficiency or leakage. Golle et al.~\cite{acns04-golle} laid foundational work, inspiring advancements in boolean queries~\cite{crypto13-oxt,sp14-blindseer,tdsc22-Ferreira}, dynamic updates~\cite{icde17-ibtree, vldb19-vbtree, ndss21-odxt}, and fuzzy search~\cite{tifs16-fuzzyfu}. Modern implementations of CSSE~\cite{crypto13-oxt, ccs18-hxt, ndss21-odxt, ccs24-doris} predominantly build upon the OXT~\cite{crypto13-oxt} framework, with first retrieving documents via the least frequent keyword (\texttt{s-term}) to minimize search complexity, then filtering for full conjunctions. This wide-used construction reduces leakage but still reveals structural information, inspiring attacks exploiting such leaks.

\noindent \textbf{Leakage-abuse attacks (LAAs). }
LAAs have emerged as a critically concerning threat, targeting the security vulnerabilities of Searchable Symmetric Encryption (SSE) systems during their real-world deployment. The first pioneering work in this area was introduced by Islam et al.~\cite{ndss12-ikk} and later improved by Cash et al.~\cite{ccs15-count}. Since then, a substantial body of related research has emerged under different adversarial models. 
Recently, a wide range of LAAs target the single-keyword search scenario. These include active attacks~\cite{usenix16-all, ndss20-revisiting, usenix23-injection}, which attempt to influence the system by injecting specific files to manipulate the search process, and passive attacks which rely on statistical analysis and correlations between leaked information and background knowledge in the form of known datasets~\cite{ndss12-ikk, ccs15-count, ndss20-revisiting, ccs21-leap, ccs23-dynamic, tifs19-ning, tifs21-xulei} or similar datasets~\cite{is14-freq, usenix21-sap, usenix21-refscore, usenix22-ihop, sp23-rethinking, ccs23-dynamic, usenix24-jigsaw} to compromise query privacy. 
Although LAAs have been extensively researched and explored in the single-keyword search scenario, a major limitation remains: single-keyword search is not practical for real-world applications. 

Zhang et al.~\cite{usenix16-all} extended their attack to the conjunctive query scenario, however, their approach relies on active file injection, which is infeasible in most real-world scenarios and incompatible with forward-secure schemes. Dijkslag et al.~\cite{c22-ckws} first explored the passive query recovery attack against secure conjunctive keyword search schemes. They proposed an easy and generic extension strategy that adapts query-recovery attacks from single-keyword searches by simply substituting the single-keyword set with a keyword conjunction set. Unfortunately, their experimental results show that the attack performs poorly on similar datasets with huge time and space overhead, even if they have access to a set of known queries as part of the attacker's knowledge. 

\noindent \textbf{Leakage suppression. }
To mitigate leakage-abuse attacks (LAAs), various defenses~\cite{ndss12-ikk, ccs15-count, infocom18-clrz, usenix20-seal} have been proposed. Obfuscation~\cite{infocom18-clrz} is a widely adopted approach, where a document matching a queried keyword is returned with probability $p$ (the true positive rate, TPR), while a non-matching document is returned with probability $q$ (the false positive rate, FPR), thereby introducing uncertainty in query outcomes. Another strategy involves volume padding. Cash et al.~\cite{ccs15-count} introduced a foundational padding technique that adjusts the volume to align with the closest multiple of a predefined integer $k$. Demertzis et al.~\cite{usenix20-seal} developed SEAL, which further modifies the response size to the nearest power of an integer $x$.

\section{Conclusion}
\label{sec:conclusion}

In this paper, we revisited efficient conjunctive SSE (CSSE) schemes and analyzed their vulnerabilities to leakage-abuse attacks (LAAs). Our investigation characterized leakage profiles of OXT-based schemes, introducing \texttt{s-term}-related leakage patterns to LAAs and discovering a novel \texttt{s-term} combination pattern. Building on these, we proposed {\sffamily S-Leak}, a three-stage passive query recovery attack. Empirical evaluations on real-world datasets validate {\sffamily S-Leak} effectiveness across diverse CSSE configurations. Our findings underscore the need to revisit security-efficiency balances in modern CSSE designs. For future work, we aim to deepen exploration of CSSE leakage patterns to enhance attack precision and develop lightweight and deployable defenses that balance security with system efficiency.



\section*{Statement on Artifacts}
To ensure the reproducibility and transparency of our research, all artifacts of this work including datasets, codes and configuration files will be released on GitHub upon the paper acceptance.

\bibliographystyle{ACM-Reference-Format}
\bibliography{ref}

\appendix

\section{Auxiliary Frequency Processing for Attack Model}
\label{app:freq_process}

\begin{figure*}[t!]
\centering
\includegraphics[width=0.95\textwidth]{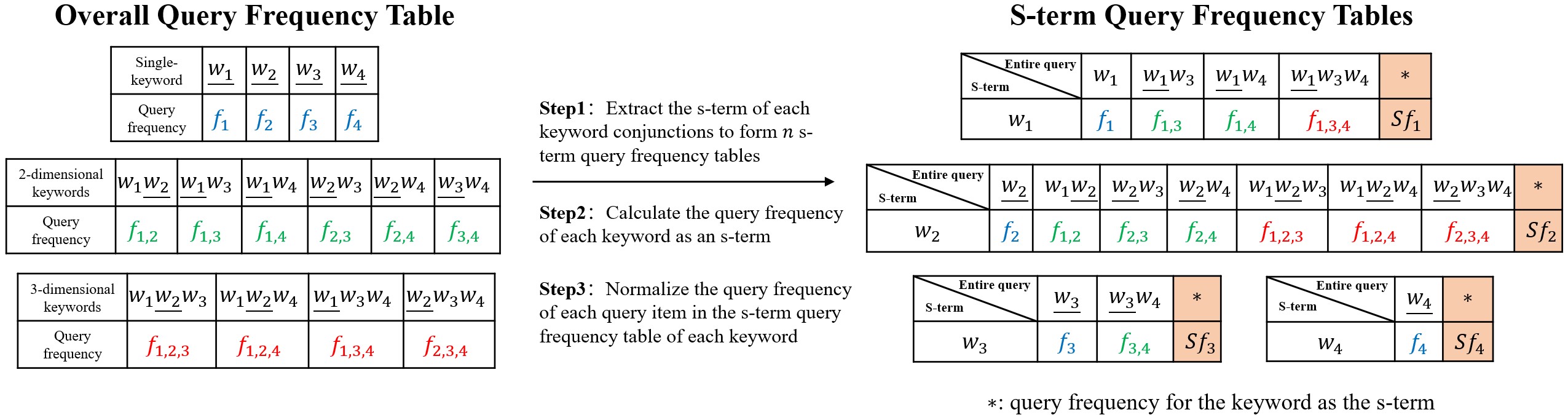}
\caption{\texttt{s-term} query frequency processing illustration with d=3 and hybrid query setting.(The underlined keyword is \texttt{s-term}.)}
\label{fig:1}
\end{figure*}
Due to the practical limitation that obtaining conjunctive query frequencies grows exponentially with the dimension of conjunctive queries, we assume that the attacker only knows the query frequencies of single-keyword queries and 2-dimensional queries. These frequencies have an inclusion relationship (e.g., the query frequency of the conjunctive query $(w_1,w_2)$ is included within the single-keyword query frequency of $w_1$). Therefore, the attacker can approximate the frequencies of queries with higher dimension by utilizing frequency estimation methods, and detailed description can be obtained in the Appendix \ref{app:freq_approx}. However, query frequencies obtained from public information do not directly provide the query frequency of keywords as \texttt{s-term}, which means that there is no readily available frequency knowledge to match the \texttt{s-term} equality pattern in the query recovery attack. Specifically, the frequency of a single-keyword query is not equal to the frequency of that keyword serving as an \texttt{s-term} in a conjunctive query. 
The frequency of a keyword acting as the \texttt{s-term} depends on many factors, including the single-keyword query frequency, conjunctive-keyword query frequency, the document frequency of the keyword in the dataset, and the probability of query varying dimension of conjunctive keywords in the hybrid query setting.

We assume the universe of keywords is $\Delta_k=[w_1,w_2,w_3,w_4]$, with the number of matching documents for each keyword being $|D(w_1)|=~3$, $|D(w_2)|=1$, $|D(w_3)|=5$, and $|D(w_4)|=6$. The attacker has obtained the single-keyword query frequency information as $f(w_1)=~0.1$, $f(w_2)=0.2$, $f(w_3)=0.3$, and $f(w_4)=0.4$. Assuming a conjunctive query scheme supporting hybrid query with the maximum dimension of conjunctive queries $d=3$, we consider the following example queries (To simplify the presentation, we list the query with keyword conjunctions.): 
$\{(\underline{w_4}),(\underline{w_1},w_3),(\underline{w_2},w_4),(\underline{w_2},w_3),(\underline{w_3},w_4), (\underline{w_2},w_3,w_4),(\underline{w_1},w_4),\\(\underline{w_3},w_4),(\underline{w_2},w_4),(\underline{w_3},w_4)\}$, where the underlined keyword in each query is the \texttt{s-term}. These queries satisfy the given single-keyword query frequencies. Now, by calculating the query frequency for each keyword as the \texttt{s-term}, we obtain $Sf(w_1)=0.2$, $Sf(w_2)=0.4$, $Sf(w_3)=0.3$ and $Sf(w_4)=0.1$, where $Sf(\cdot)$ represent the query frequency of keyword as \texttt{s-term}. It is evident that this frequency differs from the single-keyword query frequency. This discrepancy arises because the frequency of a keyword acting as the \texttt{s-term} depends on many factors, including the single-keyword query frequency, conjunctive-keyword query frequency, the document frequency of the keyword in the dataset, and the probability of query varying numbers of keywords in the hybrid query setting (In this example, $P_d(n_{search}=1)=0.1$, $P_d(n_{search}=2)=0.8$, $P_d(n_{search}=3)=0.1$.).

To this end, we process the original frequency information owned by the attacker, and construct frequency information that can match the \texttt{s-term} query frequency by following the logic of conjunctive keywords search process. The specific steps are as follows. Firstly, we combine all the query frequency information that the attacker possesses into an \textit{overall query frequency table}. For each keyword conjunction, we extract the \texttt{s-term} according to the document frequency of the corresponding keywords in the auxiliary dataset $D_a$, partitioning the overall query frequency table into $n$ \textit{\texttt{s-term} query frequency tables}. Secondly, we sum the corresponding entries in the overall query frequency table to obtain the query frequency for each keyword as the \texttt{s-term}. Next, we calculate the normalized query frequency of each entry within each \texttt{s-term} query frequency table. Specifically, we divide the frequency in the overall query frequency table into the \texttt{s-term} query frequency tables and replace the corresponding entries in the \texttt{s-term} query frequency table with the normalized frequencies. The normalized query frequency reflects the probability that, given a keyword is the \texttt{s-term} of a particular query, the query corresponds to a specific entry in the table. At this point, the attacker obtains the \texttt{s-term} query frequencies of keywords $\widetilde{\textbf{Sf}}=[\widetilde{Sf}_1,\widetilde{Sf}_2,\ldots,\widetilde{Sf}_n]$, along with the query frequencies for the conjunctive queries when $w_i$ is the \texttt{s-term}, denoted as $\widetilde{\textbf{f}_i}=[\widetilde{f_i}_1,\widetilde{f_i}_2,\ldots,\widetilde{f_i}_{\tilde{m}_i}]$. An illustration of this process is shown in Figure \ref{fig:1}, and the keyword with underline is the \texttt{s-term} of the conjunctive query.

\section{Frequency Approximation Method} 
\label{app:freq_approx}
\begin{figure*}[htbp]
    \begin{subfigure}[b]{0.35\textwidth}
        \label{fig:lucene_rho_d2_hybrid} 
        \includegraphics[width=\textwidth]{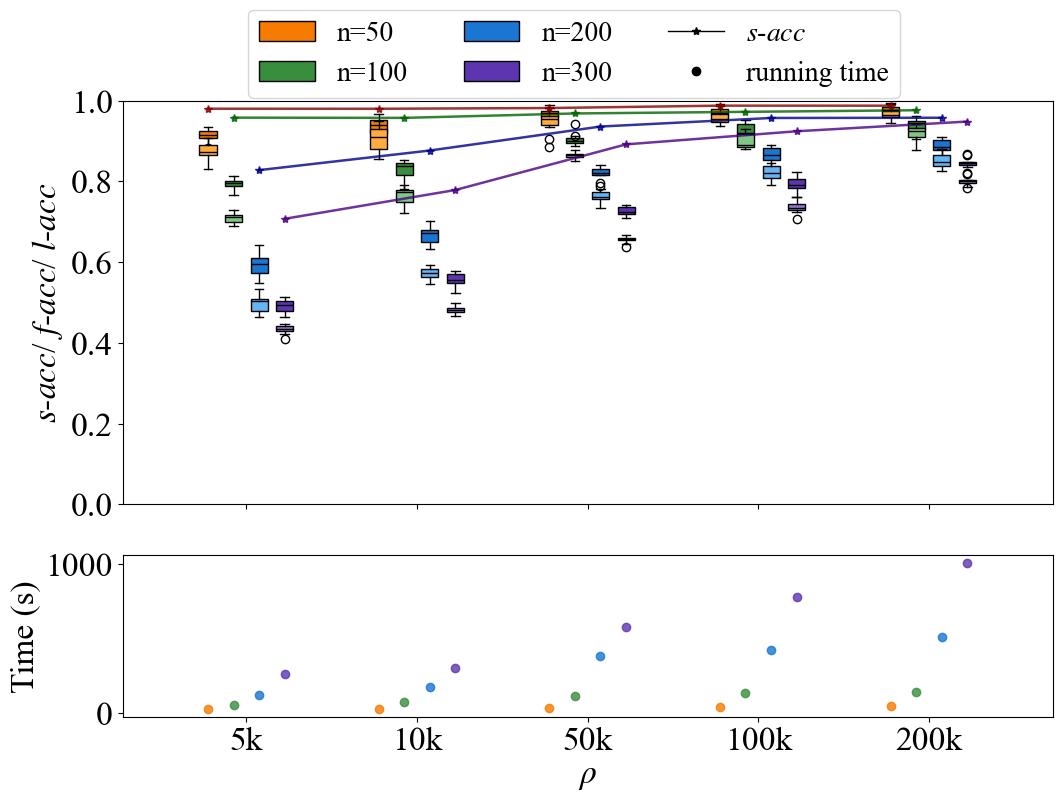}
        \caption{\normalfont Lucene $d=2$}
    \end{subfigure}
    \hspace{30pt}
    \begin{subfigure}[b]{0.35\textwidth}
        \label{fig:lucene_rho_d3_hybrid} 
        \includegraphics[width=\textwidth]{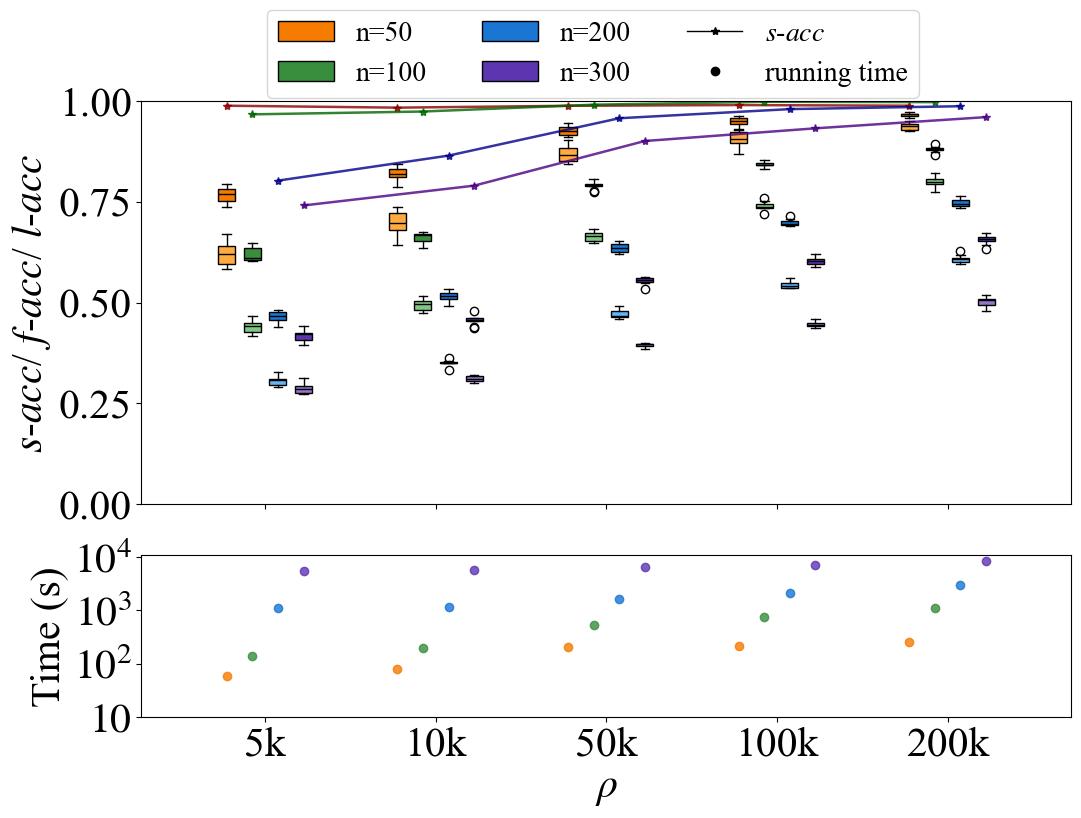}
        \caption{\normalfont Lucene $d=3$}
    \end{subfigure}
    \caption{Performance of {\sffamily S-Leak} using similar-data with hybrid queries on Lucene dataset.} 
    \label{fig:res_rho_hybrid_lucene} 
\end{figure*}

\begin{figure*}[htbp]
    \begin{subfigure}[b]{0.35\textwidth}
        \label{fig:Pd_d2}
        \includegraphics[width=\textwidth]{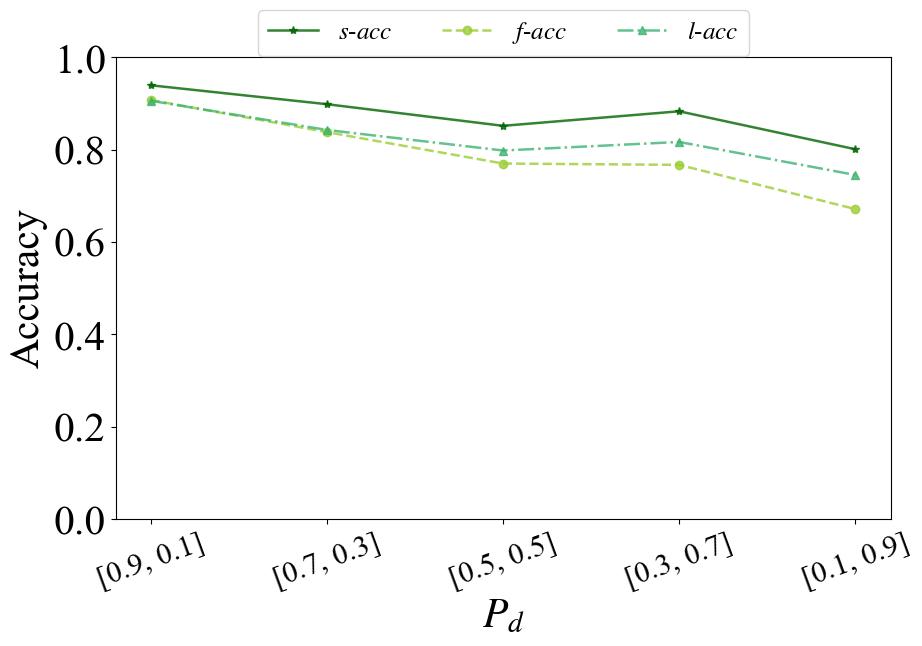}
        \caption{\normalfont Enron $d=2$}
    \end{subfigure}
    \hspace{30pt}
    \begin{subfigure}[b]{0.35\textwidth}
        \label{fig:Pd_d3}
        \includegraphics[width=\textwidth]{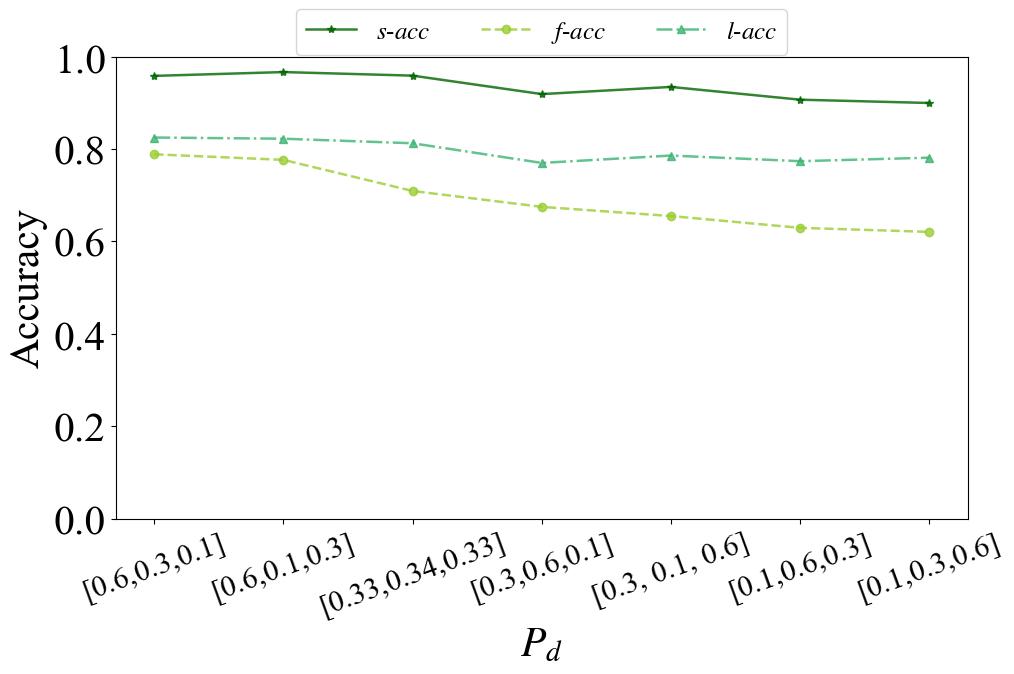}
        \caption{\normalfont Enron $d=3$}
    \end{subfigure}
    \caption{Performance of {\sffamily S-Leak} with under varying $P_d$ on Enron dataset.}
    \label{fig:res_Pd_same}
\end{figure*}
Due to the lack of complete knowledge of query frequencies, it is not possible to directly model the query frequency of \texttt{s-term}s when $d>2$. Therefore, we approximate the query frequency using an estimation approach. Specifically, we apply the \textbf{conditional independence assumption}~\cite{stoc18-conindepend} to estimate the query frequency of high-dimensional conjunctive queries based on the query frequency of individual keywords and 2-dimensional keyword conjunctions. We chose to approximate the frequency using the conditional independence assumption because it achieves reasonable accuracy at a much faster computation speed. In contrast, Monte Carlo simulations would require over 10,000 iterations to achieve similar accuracy. Estimating the query frequency of 3-dimensional conjunctive queries for 300 keyword sets over 260 weeks using Monte Carlo simulations on our laptop would take approximately 21 days, making it computationally prohibitive. 

The basic principle of the conditional independence assumption is that given two events, a third event is conditionally independent of them.

Let:

$A$ = \{client queries keyword $w_1$\},

$B$ = \{client queries keyword $w_2$\},

$C$ = \{client queries keyword $w_3$\},

$D$ = \{client queries keyword $w_4$\},

$E$ = \{client queries keyword $w_5$\}.

The attacker possesses the following frequency knowledge: (1) Query frequency of individual keywords: $P(A), P(B), P(C), P(D), P(E)$; (2) Query frequency of 2-dimensional keyword conjunctions: $P(A\cap B), P(A\cap C), P(A\cap D), P(A\cap E), P(B\cap C), P(B\cap D), P(B\cap E), P(C\cap D), P(C\cap E), P(D\cap E)$.

Using the conditional independence assumption, the query frequency of 3-dimensional keyword conjunctions can be approximated. For example, if $A$ is conditionally independent given $B$ and $C$, then $P(A\cap B \cap C) \approx P(A|B)\cdot P(A|C) \cdot P(B\cap C)$. Similarly, we can derive $P(A\cap B \cap C) \approx P(B|A)\cdot P(B|C) \cdot P(A\cap C)$, and $P(A\cap B \cap C) \approx P(C|A)\cdot P(C|B) \cdot P(A\cap B)$. The final estimated frequency is the average of these three approximations. 

This method can be extended to 4-dimensional and 5-dimensional keyword conjunctions. For example, for four keywords, $P(A\cap B \cap C \cap D) \approx P(C\cap D|A)\cdot P(C\cap D|B) \cdot P(A\cap B)$, and for five keywords, $P(A\cap B \cap C \cap D\cap E) \approx P(C\cap D\cap E|A)\cdot P(C\cap D \cap E|B) \cdot P(A\cap B)$

We evaluate the effectiveness of our frequency approximation algorithm by randomly selecting 20 keywords from the top 300 keywords in the Enron dataset and obtaining their 3-dimensional query frequency from Google Trends as ground truth. We calculate the mean squared error (MSE) between the approximated frequency matrix and the ground truth matrix over 260 weeks, which is $\text{MSE} = 3.9978 \times 10^{-5}$. It is worth noting that this approach disregards certain degree of query correlation. Consequently, the accuracy of the approximation deteriorates as the number of keywords in the conjunction increases.

\section{Adaptations to Defenses}
\label{app:adaptations}
Countermeasures such as padding and obfuscation appear to overlook the protection of associated parameters. If an attacker gains access to these parameters, they can adjust similar data to undermine the defenses, reducing the impact of padding and obfuscation on query recovery. Our specific adaptations are as follows. 
\begin{itemize}
\item[$\bullet$] \textit{Padding in SEAL}~\cite{usenix20-seal}. To adapt our attack against SEAL, the same padding method is applied to the auxiliary dataset, and the padded dataset is utilized to replace the original auxiliary dataset. When auxiliary dataset differs in size from user's dataset, we expand the auxiliary dataset to match the scale of the user dataset by data duplication, thereby preserving its original size distribution. Subsequently, the adjusted auxiliary dataset is employed to participate in the subsequent padding adaptation process.
\item[$\bullet$] \textit{Obfuscation in CLRZ}~\cite{infocom18-clrz}. Our adaptation consists of two phases, corresponding to the recovery of \texttt{s-term}s and entire queries. For the recovery of \texttt{s-term}s, after applying CLRZ, the probability that a document contains an \texttt{s-term} $w_i$ is 
\begin{equation}
\tilde{v}_i\cdot\text{TPR}+(1-\tilde{v}_i)\cdot\text{FPR}.
\label{adp:clrzm1}
\end{equation}
And for the recovery of entire queries, CLRZ does not account for the correlation between injected keywords within a document, assuming independent retention or removal of each keyword. Consequently, for a $d$-dimensional keyword conjunction, a document that originally contains this conjunction retains it with probability $\text{TPR}^{d}$, while it is removed with probability $(1-\text{TPR}^{d})$ (since the removal of any keyword in the conjunction eliminates the entire conjunction). Conversely, a document that does not originally contain the conjunction has a probability of $\text{FPR}^{d}$ of falsely including it and a probability of $(1-\text{FPR}^{d})$ of remaining unaffected. 
Recall that ${\widetilde{\textbf{V}_i}}_{g,g}^{'}$, is an estimation of the probability that a document has both keyword conjunctions ${\xi_{i}}_g$ and ${\xi_{i}}_{g'}$. Let ${\widetilde{\textbf{V}_i}}_{g,g}^{'not}$ be an estimation of the probability that a document has neither keyword conjunctions ${\xi_{i}}_g$ and ${\xi_{i}}_{g'}$. Then, the $g$,$g'$-th entry of ${\widehat{\textbf{V}_i}}$ is 
\begin{equation}
\footnotesize
({\widehat{\textbf{V}_i}})_{g,g'}=
\begin{cases}
        {\text{TPR}^{2d}} \cdot (\widetilde{\mathbf{V}_i})_{g,g'}^{'} + {\text{FPR}^{2d}} \cdot (\widetilde{\mathbf{V}_i}^{'\text{NOT}})_{g,g'} \\
        \quad + {\text{TPR}^{d}} \cdot {\text{FPR}^{d}} \cdot [1 - (\widetilde{\mathbf{V}_i})_{g,g'}^{'} - (\widetilde{\mathbf{V}_i}^{'\text{NOT}})_{g,g'}], & g \neq g', \\[10pt]
        {\text{TPR}^{d}} \cdot (\widetilde{\mathbf{V}_i}^{'})_{g,g'} + {\text{FPR}^{d}} \cdot (\widetilde{\mathbf{V}_i}^{'\text{NOT}})_{g,g'}, & g = g'.
    \end{cases}
\label{adp:clrzm3}
\end{equation}
To adapt our attack against CLRZ, the attacker simply replace $\tilde{v}_i$ in (\ref{m1:costv}) by (\ref{adp:clrzm1}) and ${\widetilde{\textbf{V}_i}}_{g,g}^{'}$ in (\ref{m3:Vu1})(\ref{m3:Vu2}) by (\ref{adp:clrzm3}).
\end{itemize}

\section{Additional Experiment Result}
\label{app:add_result}
The results of hybrid query setting on Lucene dataset are shown in \ref{fig:res_rho_hybrid_lucene}. The same as the separate query setting, our attack achieves superior performance on Lucene dataset. 

In hybrid query setting, to further mimic real-world heterogeneity, we vary $P_d$ (proportion of $d$-dimensional queries) in the hybrid setting. For $n=100$ and $\rho=100,000$, we test various $P_d$ and report $s\mbox{-}acc$, $f\mbox{-}acc$, and $l\mbox{-}acc$, as shown in Figure \ref{fig:res_Pd_same}. The general trend indicates that a higher proportion of conjunctive queries with higher dimension leads to lower attack accuracy, which is consistent with intuition. 


\end{document}